\documentclass[superscriptaddress,nofootinbib]{revtex4-1}
\usepackage[utf8]{inputenc}
\usepackage{bm}

\usepackage[dvipsnames]{xcolor}
\definecolor{red}{rgb}{0.9, 0,0}
\definecolor{cerulean}{rgb}{0., 0.42,0.9}
\definecolor{navy}{rgb}{0.05, 0.05,0.8}

\usepackage[colorlinks]{hyperref}
\hypersetup{
   colorlinks  = true,
   citecolor  = red,
	linkcolor = navy
}

\usepackage{bbm}
\usepackage{amsfonts}
\usepackage{mathrsfs}
\usepackage{bbold}
\usepackage{amsmath,amssymb}
\usepackage{subcaption}
\usepackage{siunitx}
\usepackage{slashed}

\usepackage{epsfig}
\usepackage{graphicx}     
\usepackage{url}
\usepackage{float}
\usepackage{soul}
\usepackage{pstricks}
\usepackage{color}
\usepackage{multirow}

\newcommand{\be}{\begin{equation}}
\newcommand{\ee}{\end{equation}}
\newcommand{\bea}{\begin{eqnarray}}
\newcommand{\eea}{\end{eqnarray}}
\newcommand{\beq}{\begin{eqnarray}}
\newcommand{\eeq}{\end{eqnarray}}
\def\bit{\begin{itemize}}
\def\eit{\end{itemize}}
\def\ben{\begin{enumerate}}
\def\een{\end{enumerate}}

\newcommand{\units}[1]{\mathrm{\; #1}}

\newcommand{\OO}{\mathcal{O}}

\newcommand\DN[1][\relax]{%
\ifx\relax#1\relax\else{}^{#1}\fi \!X}

\DeclareMathAlphabet\mathbfcal{OMS}{cmsy}{b}{n}

\newcommand{\eV}{\text{eV}}
\newcommand{\x}[1]{\ensuremath{\textnormal{#1}}}

\newcommand{\rp}{\right)}
\newcommand{\lp}{\left(}
\newcommand{\rb}{\right]}
\newcommand{\lb}{\left[}

\newcommand{\voma}{\vec{\omega}_a}
\newcommand{\moma}{\omega_a}
\newcommand{\omdm}{\omega_\x{dm}}
\newcommand{\omsm}{\omega_\x{sm}}
\newcommand{\tbunch}{t_\x{bunch}}

\newcommand{\trun}{t_\x{run} }
\newcommand{\mdm}{m_\x{dm}}
\newcommand{\gmt}{g_{\mu-\tau}}

\definecolor{cerulean}{rgb}{0., 0.52,0.65}

\graphicspath{{plots/}}

\begin{document}

\title{Muon g-2 and EDM experiments as muonic dark matter detectors}

\author{Ryan Janish}
\affiliation{Department of Physics, University of California, Berkeley, CA 94720}
\author{Harikrishnan Ramani}
\affiliation{Department of Physics, University of California, Berkeley, CA 94720}
\affiliation{Theoretical Physics Group, Lawrence Berkeley National Laboratory, Berkeley, CA 94720}

\begin{abstract} 
The detection of ultralight dark matter through interactions with nucleons, electrons, and photons has been explored in depth. In this work we propose to use precision muon experiments, specifically muon g-2 and electric dipole moment measurements, to detect ultralight dark matter that couples predominantly to muons. 
We set direct, terrestrial limits on DM-muon interactions using existing g-2 data, and show that a time-resolved reanalysis of ongoing and upcoming precession experiments will be sensitive to dark matter signals.   
Intriguingly, we also find that the current muon g-2 anomaly can be explained by a spin torque applied to muons from a pseudoscalar dark matter background that induces an oscillating electric dipole moment for the muon.
This explanation may be verified by a time-resolved reanalysis.
\end{abstract}

\maketitle
\tableofcontents


\section{Introduction}
Despite the presence of dark matter (DM) and its gravitational interactions being well established, its particle nature and non-gravitational interactions with the standard model (SM) are yet to be illuminated. While the elementary dark matter mass could span many orders of magnitude, the ultralight dark matter regime, $10^{-22} \textrm{eV} \le m_{\rm dm} \ll \textrm{eV}$, has received much attention recently. These ultralight particles arise naturally in solutions to tuning problems, e.g.~the axion \cite{Peccei:1977hh} and the relaxion~\cite{Graham:2015cka}, as well as in the string landscape. Furthermore, they also have attractive production mechanisms --- misalignment for scalars\cite{Preskill:1982cy}, inflationary production for vectors \cite{Graham:2015rva} and parametric resonance for both~\cite{Dror:2018pdh,Agrawal:2018vin,Co:2017mop}. 

Traditional direct detection experiments targeting the WIMP scale are not sensitive to ultralight DM, so a plethora of experiments have been performed and proposed in recent years exploiting the wave-like properties of this mass regime. Yet these have exclusively tested dark matter couplings to photons~\cite{Arvanitaki:2014faa}, electrons~\cite{Arvanitaki:2014faa}, protons, and neutrons~\cite{Graham:2013gfa,Graham:2015ifn}. Meanwhile, the muon g-2 anomaly \cite{Bennett:2006fi, Davier:2019can} has led to exploration of theories with dark forces that predominantly couple to muons and experimental proposals to find them \cite{Altmannshofer:2014pba, Chen:2017awl, Kahn:2018cqs, Escudero:2019gzq}. Similarly, dark matter itself could dominantly couple to muons. In this work, we study such models and explore the possibility of precision muon experiments directly detecting such muophilic dark matter. 

Muon g-2 and EDM experiments, such as the measurement done at BNL~\cite{Bennett:2006fi} in 2004, the ongoing work at Fermilab~\cite{Grange:2015fou} and J-PARC~\cite{Abe:2019thb}, and the proposed frozen spin experiments~\cite{Farley:2003wt, Adelmann:2006ab}~\footnote{In the final stages of this work, \cite{Graham:2020kai} appeared which primarily considers frozen spin techniques with proton storage rings to constrain pseudoscalr DM-proton wind couplings, but also briefly considers the use of muons.}, are precision efforts to track the time evolution of muon spins subject to an external magnetic field. The primary aim of the g-2 experiments~\cite{Bennett:2006fi, Grange:2015fou, Abe:2019thb} is the determination of the muon's magnetic dipole moment (MDM). However, they are sensitive to any new physics which sufficiently alters the precession dynamics of muon spins. For example, the existence of a muon electric dipole moment (EDM) has been constrained in this manner by the BNL experiment~\cite{Bennett:2008dy} and will be further tested at Fermilab and J-PARC. The frozen spin proposals are a more sensitive, dedicated search for this EDM signal. 
A coherent dark matter background may couple to muons in these experiments and alter their precession by applying a spin torque and by possibly altering their orbital trajectories. 
This results in a characteristic DM precession signal which is observable in these experiments --- we thus propose to repurpose muon precession experiments as dark matter detectors. 

DM perturbations to precession may yield a variety of signals in these experiments depending on the nature of the DM candidate. 
Some candidates would have noticeably altered the form of the precession signal in the existing analysis of BNL, allowing us to place immediate constraints.
These limits will become more stringent with ongoing and future measurements.  
In addition, some candidates may leave the form of the signal unchanged while shifting the precession frequency or amplitude. 
This is intriguing, as it provides an effective contribution to the anomalous muon MDM or the muon EDM which is set by the local DM density.
Such a DM MDM contribution may indeed explain the observed discrepancy between the BNL result and the SM prediction~\cite{Bennett:2006fi, Davier:2019can}.
Finally, an ultralight DM perturbation is generally harmonic in time, resulting in a modulation of the precession signal on timescales set by the DM mass.
The usual g-2 and EDM analysis is typically blind to this modulation as it averages over precession data spanning many DM modulation periods.
However, the modulation may be revealed with a time-resolved reanalysis of precession data.
This provides both a means of testing the background DM explanation of the muon g-2 anomaly, as well as a new opportunity for ultralight DM detection. 

The rest of this paper is organized as follows. In Sec.~\ref{sec:overview} we provide an overview of muon precession experiments. In Sec.~\ref{sec:DM-precession} we explore muon precession in the presence of a coherent dark matter field. In Sec.~\ref{sec:sensitivity} we describe the sensitivity of existing and upcoming muon precession experiments to characteristic DM signal shapes. In Sec.~\ref{sec:canidates} we consider specific DM candidates and project limits. Concluding remarks are presented in Sec.~\ref{sec:conclusion}. 


\section{Overview of Muon Spin Precession Experiments}
\label{sec:overview}

This section will provide a criminally simplified description of the physics and techniques employed to measure the precession of muon spins. 
We discuss only what is necessary to reveal the implications of these measurements on DM-muon interactions.
For more thorough reviews, see~\cite{Miller:2012opa,gorringe2015precision,Roberts:2018vsx}

\subsection{Spin Tracking via Muon Decay}
\label{sec:decay-tracking}
The spin of a muon is imprinted on the angular and energy distribution of the positrons\footnote{In this work we will refer exclusively to positive muons and their decay to positrons, while in practice experiments also employ negative muons decaying to electrons.} produced by its decay.
This is a consequence of the chiral structure of the Weak interaction. 
In the muon rest frame, the decay rate to positrons of energy $E$ emitted into a solid angle $d\Omega$ along $\hat{n}$ depends on the overlap of $\hat{n}$ with the muon spin $\vec{S}$: 
\begin{align}
\label{eq:rest-decay-spectrum}
 \frac{d\Gamma}{dE \; d\Omega} = \Gamma_0(E) 
 \lp 1 - \mathcal{A}(E) \; \hat{S} \cdot \hat{n} \rp
\end{align}
where the asymmetry factor $\mathcal{A}(E)$ is positive\footnote{The sign of $\mathcal{A}(E)$ is reversed for electrons produced by the decay of negative muons.} at the relevant energies.
The outgoing positron flux is emitted predominantly parallel to the muon spin, with the correlation becoming stronger for higher energy positrons~\cite{gorringe2015precision}.
The average spin of an ensemble of muons may thus be inferred by measuring the distribution of decay positrons. 
This technique is employed by the BNL, Fermilab, and J-PARC g-2 experiments. 
Two specific observables are measured in each experiment, a total count and a vertical count, each of which tracks a particular component of the muon spin.

\paragraph{Total Count.}
In a lab frame the highest energy decay positrons are those emitted along the muon momentum $\vec{p}$, so the lab frame energy may serve as a proxy for outgoing direction. 
As positrons are predominantly emitted parallel to the muon spin, it follows that more positrons will be produced at the highest possible energies if the muons' spin and momentum are anti-aligned than if they are aligned.
The rate of positrons emitted over all directions with a lab frame energy $E$ depends on the overlap of $\hat{S}$ and $\hat{p}$:
\begin{align}
\label{eq:lab-decay-spectrum}
  \lp \frac{d\Gamma}{dE} \rp_{\vec{p}} = 
  \Gamma_0'(E) \lp 1 - A'(E) \; \hat{S} \cdot \hat{p} \rp
\end{align}
The total count $N_T(t)$ is the number of positrons emitted above a carefully chosen energy threshold, which from~Eqn.~\eqref{eq:lab-decay-spectrum} has the form
\begin{align}
\label{eq:NE-general}
  N_T(t) \propto e^{-t/\tau_\mu} \lb \vphantom{e^{-t/\tau_\mu}} 
  1 + A \lp \vec{S} \cdot \hat{p} \rp \rb
\end{align} 
for an energy-dependent constant $A$ and the dilated muon lifetime $\tau_\mu$~\cite{Miller:2012opa}.
The time-evolution of $N_T(t)$ thus records the evolution of the projection of the muon spin along its momentum. 

\paragraph{Vertical Count.}
The second observable is the difference in the number of positrons emitted with a velocity component parallel and anti-parallel to the vertical direction, defined as the direction of the experiment's large, static magnetic field $\hat{B}$. 
From~Eqn.~\eqref{eq:rest-decay-spectrum}, this is proportional to $\hat{S}\cdot\hat{B}$ and thus probes the component of muon spin along the magnetic field. 
Instead of a differenatial count, an analgous quantity may be measured which is also proportional to the vertical component of the spin, such as the average vertical angle of outgoing positrons~\cite{Chislett:2016jau,Bennett:2008dy}.
We will refer to this measurment generically as the `vertical count' $\Delta N_B(t)$, which has the form
\begin{align}
\label{eq:DNB-general}
  \Delta N_B(t) \propto e^{-t/\tau_\mu} \lp \vphantom{e^{-t/\tau_\mu}}
  \vec{S} \cdot \hat{B} \rp.
\end{align}

\subsection{Precession Signals}
\label{sec:expected-signals}

All the muon spin precession experiments we consider, observe decaying muons which are executing cyclotron orbits in a uniform, static magnetic field $\vec{B}$. 
The muon spin precesses in $\vec{B}$ and any additional EM fields which are present. 
The experiments are designed to measure the intrinsic muon MDM and/or EDM, so we briefly describe here the expected precession signals in that case. 
This will elucidate the specific design and data analysis choices made in these experiments (see Sections~\ref{sec:exp-data-analysis} and~\ref{sec:exp-deets}), as well as introduce the notions needed to derive the DM-induced precession signals in Section~\ref{sec:canidates}.

In the lab frame, muons are held in circular orbits in a plane perpendicular to $\vec{B}$. 
They orbit with the cyclotron frequency $\vec{\omega}_C$, given by the vertical magnetic field $\vec{B}$ and possibly a radial electric field $\vec{E}$~\cite{Miller:2012opa}:
\begin{align}
\label{eq:omega-cyclotron}
  \vec{\omega}_C = - 
  \frac{q}{m} \lb \frac{1}{\gamma} \vec{B} - \lp \frac{\gamma}{\gamma^2 - 1} \rp \lp \vec{v} \times \vec{E} \rp \rb.
\end{align}
Note that for radial $\vec{E}$, $\vec{\omega}_C$ is parallel or anti-parallel to $\vec{B}$. 
We ignore for the moment non-radial $\vec{E}$ and the possibility of muons having non-zero momentum along $\vec{B}$, which would cause a deviation from circular orbits. 

It is useful to view the evolution of the muon spin in the rotating muon rest frame (RMRF). 
This is a non-inertial frame in which the muon is at rest and the velocity of the lab always points in the same direction, which we take to be the $y$-direction. 
To reach this frame at a particular time $t$, we start with a Cartesian lab frame with $\vec{B}$ in the $z$-direction, rotate so the muon momentum is in the $y$-direction, and then boost along $\hat{y}$ so the muon is at rest. 
For muons in circular, cyclotron orbits, the $z$-axis of the lab frame and RMRF coincide. 
The momentum and vertical components of $\vec{S}$ appearing in the decay counts~Eqn.~\eqref{eq:NE-general} and~Eqn.~\eqref{eq:DNB-general} are respectively the $y$ and $z$ components of spin in the RMRF. 

The muon spin $\vec{S}$ in the RMRF evolves according to a precession equation
\begin{align}
\label{eq:RMRF-eqn}
 \frac{d}{dt}\vec{S} = \vec{\omega}_a \times \vec{S}
\end{align}
where we take $t$ to be the lab time. 
The precession frequency $\vec{\omega}_a$ is given by three distinct contributions: 
\begin{align}
\label{eq:most-generl-omegea_a}
 \vec{\omega}_a = \frac{1}{\gamma} \vec{\omega}_\tau + \vec{\omega}_T - \vec{\omega}_C.
\end{align}
$\vec{\omega}_\tau$ is the result of the net torque on the muon spin in the RFMR, with the factor of $\gamma$ due to taking the derivative with respect to lab time in~Eqn.~\eqref{eq:RMRF-eqn}.
In this case $\vec{\omega}_\tau$ is due entirely to the EM fields $\vec{E'}$ and $\vec{B'}$ in that frame:
\begin{align}
  \vec{\omega}_\tau = \frac{e g_\mu}{2 m_\mu} \vec{B}' 
   + 2 d_\mu \vec{E}'
\end{align}
where $m_\mu$, $g_\mu$ are the muon mass and gyromagnetic, and $d_\mu$ is the intrinsic muon EDM. 
$\vec{\omega}_T$ is the Thomas precession, arising from the accelerated motion of the muon. 
This may be computed in terms of the lab frame trajectory $\vec{v}(t)$ of the muon~\cite{eddington1924mathematical}:
\begin{align} 
  \vec{\omega}_T = \lp \frac{\gamma^2}{\gamma + 1} \rp
  \frac{d\vec{v}}{dt} \times \vec{v}
\end{align}
Finally, $\vec{\omega}_C$ is the cyclotron frequency~Eqn.~\eqref{eq:omega-cyclotron}, which appears because the RMRF rotates at $\vec{\omega}_C$ relative to the lab.
All of these contributions may be expressed in terms of the lab frame fields $\vec{E}$ and $\vec{B}$, which yields
\begin{align}
\label{eq:omega_a}
 \vec{\omega}_a = 
 -\frac{e}{m_\mu} a_\mu \vec{B} + 
 \frac{e}{m} & a_\mu \lp \frac{\gamma}{\gamma + 1} \rp 
   \lp \vec{B} \cdot \vec{v} \rp \vec{v} +
 \frac{e}{m_\mu} \lp a_\mu - \frac{1}{\gamma^2 - 1}
  \rp \vec{v} \times \vec{E} -
 2\,d_\mu \lp \vec{E} + \vec{v} \times \vec{B} \rp
\end{align}
where $a_\mu = g_\mu/2 -1$. 
Note that the $\vec{v} \cdot \vec{B}$ term vanishes for circular orbits.
The spin trajectory in the RMRF is uniform, circular precession with angular velocity $\vec{\omega}_a$, since $\vec{\omega}_a$ is time-independent in that frame. 

We take the muon spin to be initially parallel or anti-parallel to the momentum, as is the case in the experiments considered.~\footnote{
BNL and Fermilab always have this configuration. J-PARC will have the ability to rotate the spin before placing the muons onto cyclotron orbits~\cite{Artikova:2017xky}, but for simplicity we focus here on the parallel configuration.}
The g-2 experiments are designed so that the first term in Eqn.~\eqref{eq:omega_a} dominates. 
And with the simplifying assumption of vanishing EDM, $\vec{\omega}_a$ is in the $z$-direction so the spin precesses in the $xy$-plane. 
The vertical component is zero and the momentum component is harmonic:
\begin{align}
\label{eq:expected-Sp-mdm}
  S_y &= S \cos\lp\omega_a t\rp \\
\label{eq:expected-SB-mdm}
  S_z &= 0
\end{align}
where the oscillation frequency is the magnitude $\omega_a = |\vec{\omega}_a|$.
For a small but nonzero $d_\mu$, $\vec{\omega}_a$ is slightly tilted in the RMRF from the $z$-direction into the $x$-direction, by an angle proportional to $d_\mu$. 
The spin now precesses in a plane slightly tilted from the $xy$-plane and has a harmonic vertical component in addition the harmonic momentum component:
\begin{align}
\label{eq:expected-Sp-mdm+edm}
  S_y &\approx S \cos\lp\omega_a t\rp \\
\label{eq:expected-SB-mdm+edm}
  S_z &\propto d_\mu \sin\lp\omega_a t\rp
\end{align}
We may therefore think of the total count~Eqn.~\eqref{eq:NE-general} as probing the precession magnitude $|\vec{\omega}_a|$ and the vertical count in ~Eqn.~\eqref{eq:DNB-general} as probing components of $\vec{\omega}_a$ which are perpendicular to $\vec{B}$.  
Note that a nonzero EDM always increases the magnitude of $\omega_a$ (see Eqn.~\eqref{eq:omega_a}). 
However, from the total count alone this is indistinguishable from the muon having zero EDM and an anomalous gyromagnetic ratio instead~\cite{Feng:2001mq}.
Breaking this degeneracy is a key motivation for the vertical count~\cite{Bennett:2008dy}. 

The g-2 experiments allow a simultaneous measurement of $a_\mu$ and $d_\mu$. 
However, better sensitivity to $d_\mu$ can be achieved with a dedicated search. 
One example is the frozen spin technique, in which the experiment is designed so that all of the terms in Eqn.~\eqref{eq:omega_a} cancel except for the $d_\mu$ term. 
Precession is then entirely due to an EDM, and the expected trajectory is
\begin{align}
\label{eq:expected-Sp-frozen}
  S_y &\approx S \cos\lp\omega_a t\rp \\
\label{eq:expected-SB-frozen}
  S_z &\approx S \sin\lp\omega_a t\rp.
\end{align}
Note that the amplitude of the vertical component is no longer suppressed by $d_e$ and now $\omega_a \propto d_e$. 
A measurement of the EDM can now be made by determining $\omega_a$ from the vertical count. 

\subsection{Data Analysis}
\label{sec:exp-data-analysis}

We consider first the analysis of the g-2 experiments. 
The anticipated uniform precession of~Eqn.~\eqref{eq:expected-Sp-mdm+edm} and~Eqn.~\eqref{eq:expected-SB-mdm+edm} would yield total and vertical counts in the form of decaying harmonic oscillations, 
\begin{align}
\label{eq:NE-constantomega-stack}
  N_T(t) &\propto \, e^{-t/\tau_\mu} \lb \vphantom{e^{-t/\tau_\mu}} 
  1 + A \cos\lp \moma t + \phi \rp \rb \\
\label{eq:DNB-constantomega-stack}
  \Delta N_B(t) &\propto \, e^{-t/\tau_\mu} \, d_\mu \, \sin\lp \moma t + \phi \rp.
 \end{align}
These signals are observed from a succession of muon bunches, with the number of bunches ranging from $10^6$ to $10^8$ and occurring over the course of years-long experimental run times (see Section~\ref{sec:exp-deets}).
The time-series of positron counts for every individual bunch are recorded and timestamped with GPS timing~\cite{Carey:2009zzb}. 

The experiments seek to extract from the ensemble of single-bunch signals an estimate of $\omega_a$ and $d_\mu$. 
Since these quantities are expected to be constant in time, a sensible technique is to align and sum the signals from each bunch, creating a stacked signal with a large signal-to-noise ratio (SNR). 
The alignment may be readily done with the total count, which has $\text{SNR} > 1$ even within each bunch~\cite{Bennett:2006fi}.
But this cannot be done independently with the vertical counts, as the expected SM amplitude is much smaller than the noise.
However, as the vertical count oscillation for an EDM has a fixed phase shift relative to the total count oscillation (see Eqns.~\eqref{eq:NE-constantomega-stack}~and~\eqref{eq:DNB-constantomega-stack}), the same alignment shifts used in the total count may be used to coherently sum the vertical count~\cite{Bennett:2008dy}. 
The two resulting stacked signals may then be fit to deduce $\omega_a$ and $d_\mu$. 

Stacking of the vertical counts may also be used in frozen spin experiments. 
In that case $\omega_a$ is small, being proportional to $d_e$, and only the leading-order behavior of Eqn.~\eqref{eq:expected-SB-frozen} is observed, $S_z \approx S \, \omega_a t$. 
Alignment is therefore not an issue, and the vertical counts may be summed and then fit for the slope $\omega_a$, which determines $d_e$. 

\subsection{Specific g-2 and EDM Experiments}
\label{sec:exp-deets}

While the BNL, Fermilab, J-PARC, and frozen spin experiments all follow the general strategy outlined in Sections~\ref{sec:decay-tracking} and~\ref{sec:expected-signals}, they differ in their detailed implementation. 
We outline here the differences which are relevant to the detection of DM precession. 
Unless otherwise cited, the specific values used here are taken from the experimental documentation~\cite{Bennett:2006fi,Grange:2015fou,Abe:2019thb,Adelmann:2006ab}.

\paragraph{BNL.}
Muons were held on their cyclotron orbits with an additional electric field $\vec{E}$, configured as a Penning trap. 
This field is radial in the plane of the orbit, as in~Eqn.~\eqref{eq:omega-cyclotron}, and yields a vertical restoring force above and below the orbital plane. 
To minimize the need to carefully measure $\vec{E}$, the muon momentum is chosen such that the $\vec{v} \times \vec{E}$ term in the precession frequency $\vec{\omega}_a$ in~Eqn.~\eqref{eq:omega_a} vanishes. 
The boost factor of these muons is known as the \emph{magic gamma}, $\gamma_\text{magic} \approx 29.3$. 
This also removes any energy-dependence from $\omega_a$, which is now determined only by the magnetic field.
A field $\vec{B} \approx 1.7 \units{T}$ was used, which yields a SM precession period $2\pi/\omega_a \approx 4 \units{\mu s}$.
Decay positrons were collected by 24 calorimeters stations located along the inner radius of the muon orbit.

Muon precession is observed in a succession of muon bunches. 
Each bunch produced an oscillatory decay signal of duration $660 \units{\mu s}$, which is roughly ten muon lifetimes at $\gamma_\text{magic}$ and contained about $150$ spin precession periods.
Each data run lasted around $5 \units{months}$, observing roughly $10^6$ bunches and $10^9$ decay positrons in total. 
There runs were completed in three consecutive years, from 1999 to 2001, which measured $a_\mu$ to a precision of $0.5 \units{ppm}$ and found a $3.3\sigma$ discrepancy from the SM prediction~\cite{Keshavarzi:2019abf, Davier:2019can}. 
Note that this experiment directly measured $\omega_a$ in~Eqn.~\eqref{eq:omega_a}, and a determination of $a_\mu$ requires an independent measurement of the muon mass. 
This was taken from measurements of the hyperfine splitting of muonium performed a few years earlier at LAMPF~\cite{liu1999high}.

Three different observables were used to obtain a vertical count~\cite{Bennett:2008dy}. 
The least systematically difficult of these was the average outgoing angle of decay positrons relative to the orbital plane, which was measured with a tracking detector placed in front of one calorimeter station.
Fewer positrons were therefore detected in this count than in the total count.
This allowed a limit to be set on the muon EDM; $|d_\mu| < 1.9 \cdot 10^{-19} \units{e \cdot cm}$.
Converting this into a relative precision for measuring the perpendicular, EDM-induced component of $\vec{\omega}_a$, we have $\delta \omega_{EDM}/\omega_a \approx 0.5 \cdot 10^{-3}$.

\paragraph{Fermilab.}
The Fermilab measurement is very similar to that of BNL, seeking to improve primarily by increased statistics.
It employs a Penning trap electric field and uses muons at $\gamma_\text{magic}$. 
The static field is slightly smaller, $\vec{B} \approx 1.45 \units{T}$.
Decay positrons are counted with 24 calorimeter stations along the inner orbit radius. 
A vertical count is made using the average positron decay angle, obtained with two tracking detectors that have significantly increased acceptance compared to that of BNL.

The bunch duration and the number of positrons detected per bunch is similar, however the average bunch cadence is increased, allowing about $10^8$ bunches and $10^{11}$ total positrons to be observed during a roughly $5 \units{month}$ run. 
This is expected to improve the precision on $a_\mu$ to $0.1 \units{ppm}$.
$a_\mu$ will be extracted from $\omega_a$ using the same LAMPF muonium measurements as BNL~\cite{liu1999high}.
The enhanced tracking detection will significantly improve the measurement of the EDM, with an expected limit of $|d_\mu| \lesssim 2 \cdot 10^{-21} \units{e \cdot cm}$ or $\delta \omega_{EDM}/\omega_a \approx 0.5 \cdot 10^{-5}$.

\paragraph{J-PARC.}
The J-PARC experiment will take a difference approach than BNL and Fermilab, seeking a measurement of $a_\mu$ and the muon EDM with qualitatively different systematics and experimental challenges.
J-PARC employs no electric field, so $\omega_a$ is again set only by the magnetic field, in this case $\vec{B} \approx 3 \units{T}$, while allowing the use of slower muons, $\gamma \approx 3$. 
The muons will be held in orbit with a weak radial magnetic field, which vanishes in the orbital plane and varies along the vertical direction.  
Detection for both the total and vertical count will be done with tracking detectors that record the spiral trajectory of decay positrons in the static magnetic field. 

The timescales involved in this approach are naturally shorted, as slower muon have a shorter dilated lifetime. 
Each bunch will last around $40 \units{\mu s}$, which is roughly $6$ muon lifetimes at $\gamma\approx 3$ and contains about $20$ spin precession periods. 
Each bunch is expected to result in about $10^3$ detected positrons, with $10^8$ bunches and $10^{11}$ positrons observed in total. 
The final precision is expected to be similar to that of Fermilab and BNL, $0.5 \units{ppm}$ on $a_\mu$ and $|d_\mu| \lesssim 2 \cdot 10^{-21} \units{e \cdot cm}$.
In addition, J-PARC is planning to perform new measurements of muonium spectroscopy using their muon source~\cite{Strasser:2019fbk} which may be used to deduce $a_\mu$ from the g-2 data. 

\paragraph{Frozen Spin EDM Experiments}
The frozen spin technique is newer than the g-2 approach, and a muon EDM search using these methods is still conceptual. 
We follow~\cite{Artikova:2017xky}, which studies the possibility of using slow muons of $\gamma \approx 1.5$ in a compact storage ring of $B\approx 1~\x{T}$. 
An applied, radial electric field is used to cancel the precession of $B$, so that $\omega_a \propto d_e$. 
With future, high-intensity muon sources, this search can reach a sensitivity of 
$|d_\mu| \lesssim 10^{-25} \units{e \cdot cm}$ or $\delta \omega_{EDM}/\omega_a \approx 10^{-9}$.
In order to estimate the sensitivity to an oscillating DM signal, we assume that such an experiment takes data over a $3~\x{year}$ timespan, with each muon bunch having a duration of about $50 \units{\mu s}$.


\section{DM Perturbed Precession}
\label{sec:DM-precession}

In this section, we consider the evolution of muon spins in a coherent, non-relativistic DM background. 
We follow the muon spin in the RMRF, defined in Section~\ref{sec:expected-signals}.
The most general equation of motion for the spin is a precession equation with a possibly time-dependent precession frequency:
\begin{align}
\label{eq:precession-eqn-general}
  \dot{\vec{S}} = \voma(t) \times \vec{S}.
\end{align}
In the g-2 experiments at BNL, Fermilab, and J-PARC the SM prediction for this frequency is constant in time and given by 
\begin{align}
\label{eq:omega_a_simplified}
 \vec{\omega}_a = 
 -\frac{e}{m_\mu} a_\mu B \, \hat{z}
\end{align}
where $B$ is the magnitude of the lab frame magnetic field, as described in Section~\ref{sec:expected-signals} and~\ref{sec:exp-deets}.
In the frozen spin proposal the SM prediction is $\voma = 0$. 
We will refer to this prediction in either case as $\vec{\omega}_\x{sm}$, the SM precession frequency. 
DM interactions may alter $\voma(t)$ by either perturbing the muon's orbital trajectory or by effecting the torque on the muon spin in the RMRF. 
In either case, the small DM perturbations may be linearized and $\omega_a(t)$ may be written as 
\begin{align}
  \voma(t) = \omsm \hat{z} + \vec{\omega}_\x{dm}(t)
\end{align}
where $\vec{\omega}_\x{dm}(t)$ is the contribution from DM-muon interactions.

The DM field value will oscillate at a frequency equal to the DM particle mass $\mdm$, and so the frequency perturbation $\vec{\omega}_\x{dm}(t)$ will similarly contain oscillatory components.
We review here the precession trajectories that result from a perturbation with a single harmonic component of frequency $m$.
Note that for a particular DM candidate, the frequency $m$ of the perturbation may not be $\mdm$ but rather a multiple of $\mdm$. 
The direction of $\vec{\omega}_\x{dm}(t)$ plays a significant role, so we consider separately parallel perturbations for which $\vec{\omega}_\x{dm}(t) = \omdm(t) \; \hat{z}$ and perpendicular perturbations for which $\vec{\omega}_\x{dm}(t) \cdot \hat{z} = 0$.

\subsection{Parallel Perturbations}
\label{sec:parallel-perturbation}

If $\vec{\omega}_\x{dm}(t) = \omdm(t) \; \hat{z}$, the precession equation
\begin{align}
	\dot{\vec{S}} = \lb \omsm + \omdm(t) \vphantom{\hat{B}} \rb \lp \hat{z} \times \vec{S} \rp
\end{align}
may be solved exactly. 
The spin precesses about $\hat{z}$ with an instantaneous angular speed $\omsm + \omdm(t)$.
A spin $\vec{S}$ which is initially parallel to the momentum and perpendicular to $\vec{B}$ precesses as 
\begin{align}
\label{eq:Sp-parallel}
	S_y(t) &= S \cos\lp \omsm t + \int_0^t dt'\; \omdm(t')\rp \\ 
 S_z(t) &= 0.
\end{align}
This may be compared to the expected SM precession with $d_\mu = 0$, given in Eqns.~\eqref{eq:expected-Sp-mdm}~and~\eqref{eq:expected-SB-mdm}.
The parallel perturbation results in a pure frequency modulation of the total count, and does not produce a signal in the vertical count. 
For a harmonic perturbation $\omdm(t) = \omdm \cos\lp m t + \alpha\rp$, this has the form
\begin{align}
\label{eq:Sp-parallel-harmonic}
 S_y(t) &= S \cos\lp \omsm t +
  \frac{\omdm}{m} \big[ \sin(m t + \alpha) - \sin(\alpha) \big] \rp \\ 
 S_z(t) &= 0.
\end{align}

\subsection{Perpendicular Perturbations}
\label{sec:perpendicular-perturbation}

Next we consider a perturbation to the precession frequency which is perpendicular to $\vec{\omega}_\x{sm}$. 
For concreteness we take this to lie in the $x$-direction of the RMRF, $ \vec{\omega}_\x{dm}(t) = \omdm(t) \; \hat{x} $, which corresponds to a precession frequency perpendicular to both $\vec{B}$ and the muon momentum, as in the case of an EDM (see Eqn~\eqref{eq:omega_a}).\footnote{The case of $\vec{\omega}_\x{dm}(t)$ parallel to the momentum ($\hat{y}$ in the RMRF) is analogous, with the only change being the value of the relative phase between the oscillation of $S_z$ and $S_y$.} 

We focus on a quasistatic perturbation, that is $\omdm(t)$ which varies at a characteristic rate $m \ll \omsm$. 
This is not true in the frozen spin setup, which we consider separately in Section~\ref{sec:trajectory-resonance-and-froze}.
Then the spin executes circular precession locally in time with a slowly-evolving instantaneous frequency $\omsm \, \hat{z} + \omdm(t) \, \hat{x}$.
The WKB solution to~Eqn.~\eqref{eq:precession-eqn-general} at leading order in $m/\omsm$ and $\omdm/\omsm$ gives:
\begin{align}
 \label{eq:Sp-perp-wkb}
 S_y(t) &\approx S \cos\lp \omsm t + 
  \frac{1}{2} \int_0^t dt'\; \frac{\omdm^2(t')}{\omsm} \rp \\	
 \label{eq:SB-perp-wkb}
 S_z(t) &\approx S \frac{\omdm(t)}{\omsm} \sin\lp \omsm t + 
  \frac{1}{2} \int_0^t dt'\; \frac{\omdm^2(t')}{\omsm} \rp,
\end{align}
for a spin initially parallel to the momentum. 
This may be compared to the expected precession with $d_\mu \neq 0$, given in Eqns.~\eqref{eq:expected-Sp-mdm}~and~\eqref{eq:expected-SB-mdm}.

The perpendicular perturbation produces a frequency modulation in the total count which scales as $\omdm^2$.
This is because the oscillation of the total count is sensitive only to the magnitude of $\voma(t)$.
The perturbation also yields a non-zero vertical count, which oscillates with a fixed phase shift relative to the total count and has an amplitude modulation which is linear in $\omdm$. 
This amplitude is independent of $m$ as it is due to the tilting of $\voma(t)$ away from $\hat{z}$, which is set by $\omdm$ alone --- taking $\omdm(t)$ to be static in~Eqn.~\eqref{eq:SB-perp-wkb} recovers the tilted precession signal of~Eqn.~\eqref{eq:expected-SB-mdm}.

For a harmonic perturbation $\omdm(t) = \omdm \cos\lp m t + \alpha\rp$, the quadratic scaling of~Eqn.~\eqref{eq:Sp-perp-wkb} produces both a net frequency shift and a frequency modulation at frequency $2m$. 
The resulting spin trajectory is 
\begin{align}
 \label{eq:Sp-perp-wkb-harmoinc}
 S_y(t) &\approx S \cos\lp \bar{\omega} t + \Phi\lb t \rb \rp \\ 
 \label{eq:SB-perp-wkb-harmoinc}
 S_z(t) &\approx S \, \frac{\omdm}{\omsm} \cos\lp m t + \alpha \rp
 \sin\lp \bar{\omega} t + \Phi\lb t \rb \rp \\ 
 \x{where:}\;\; 
  \bar{\omega} &= \omsm + \frac{1}{4} \frac{\omdm^2}{\omsm} \\
  \Phi \lb t \rb &= 
    \frac{1}{8} \frac{\omdm^2}{\omsm m} \lb 
    \vphantom{\frac{\omdm^2}{\omsm m}} \sin(2 m t + 2 \alpha) - \sin(2 \alpha) \rb 
\end{align}

\subsection{Resonance and Frozen Spin}
\label{sec:trajectory-resonance-and-froze}

The amplitude of the vertical count in the case of a perpendicular perturbation scales as $\omdm/\omsm$, as in Eqn.~\eqref{eq:SB-perp-wkb-harmoinc}.
The suppression by $\omsm$ is due to the following mechanism. 
The action of a perpendicular $\vec{\omega}_\x{dm}$ in the RMRF is to rotate the spin out of the $xy$-plane, and this rotation is either towards the $+\hat{z}$ direction or the $-\hat{z}$ direction depending on the polar angle of the spin in the $xy$-plane. 
Specifically, the spin rotates towards the direction of $\vec{\omega}_\x{dm} \times \vec{S}$. 
But the dominant motion of $\vec{S}$ is rotation in the $xy$-plane at frequency $\omsm$, and so the action of $\vec{\omega}_\x{dm}$ is not coherent --- it raises $\vec{S}$ for half of the SM period $T_\x{sm}$ and then lowers it for the next half-period. 
The maximal vertical component $S_z$ that may develop is limited by the SM rotation to be $S \, \omdm T_\x{sm} \sim S \, \omdm/\omsm$. 

This suppression is not fundamental.
It is the by-product of an experimental design optimized for the measurement of $\omsm$ itself and can be removed by using a different approach.
There are two natural possibilities for this: the frozen spin technique and resonance.
We discuss the spin trajectory in each of these cases below, focusing only on the vertical component $S_z$ as the vertical count is the most sensitive in these setups. 
Both techniques can achieve maximal coherence in the vertical signal, i.e.~an oscillation in $S_z$ with an amplitude $\sim S$. 
Indeed, they are conceptually the same technique as they both involve matching the frequency $\omsm$ to $m$, with the distinction being whether this results in $\omsm \approx 0$ or $\omsm \neq 0$. 

\paragraph{Frozen Spin}
The frozen spin technique was invented for measuring intrinsic, static EDMs~\cite{Farley:2003wt}, and is thus most sensitive to static perturbations.
In our case, this means modulation frequencies $m$ such that $m\, \tbunch \ll 1$, where $\tbunch$ is the duration of a single muon bunch. 
This method engineers $\omsm = 0$, i.e.~it \emph{freezes} the spin in the $xy$-plane (see Section~\ref{sec:exp-deets}). 
Eqns.~\eqref{eq:Sp-perp-wkb-harmoinc} and~\eqref{eq:SB-perp-wkb-harmoinc} are no longer valid in this regime, however the trajectory may be readily found as the total precession frequency in the RMRF varies only in magnitude, analogous to the parallel perturbation of Section~\ref{sec:parallel-perturbation}. 
The spin rotates about $\hat{x}$ with an instantaneous angular speed $\omega_\x{dm}(t) = \omdm \cos(m t + \alpha)$. 
This yields:
\begin{align}
\label{eq:S-frozen}
 S_y(t) &= S \cos\lp \frac{\omdm}{m} \lb \vphantom{\frac{\omdm}{m}}
  \sin(m t + \alpha) - \sin(\alpha) \rb \rp \\ 
 S_z(t) &= S \sin\lp \frac{\omdm}{m} \lb \vphantom{\frac{\omdm}{m}}
  \sin(m t + \alpha) - \sin(\alpha) \rb \rp,
\end{align}
where we have chosen $\vec{\omega}_\x{dm}$ to be along $\hat{x}$ and the spin initially along $\hat{y}$, as in Section~\ref{sec:perpendicular-perturbation}.

In the static limit, this yields a vertical signal
\begin{align} 
\label{eq:Sz-frozen-static}
 S_z(t) &\approx S \sin\lb \vphantom{S} 
 \omdm \cos\lp \alpha \rp \, t \rb, \;\;\;\; \lb m \, \tbunch \ll 1\rb
\end{align}
with no amplitude suppression. 
Note that this is a uniform rotation over a single bunch only. 
For a later bunch the value of $\alpha$ changes and the rotation frequency may have an opposite sign. 
For large $m$ the oscillation of $\omdm(t)$ introduces a new source of decoherence. 
In this case the vertical signal is 
\begin{align} 
\label{eq:Sz-frozen-nonstatic}
 S_z(t) &\approx S \, \frac{\omdm}{m} \lb \vphantom{\frac{\omdm}{m}} 
  \sin\lp m t + \alpha\rp
  - \sin\lp\alpha\rp \rb, \;\;\;\; \lb m \, \tbunch \gg 1\rb,
\end{align}
where we have assumed $m \gg \omdm$ as well, which is true in the cases we consider. 
The amplitude is now suppressed by $\omdm/m$. 
This is due to the fact that the spin's rotation about $\hat{x}$ is oscillating between clockwise and counter-clockwise motion at the DM frequency $m$, and after integrating this angular speed the vertical displacement of the spin scales as $m^{-1}$. 
This effect is analogous but physically distinct from that which produces the $\omdm/\omsm$ scaling of Eqn.~\eqref{eq:SB-perp-wkb-harmoinc}.
If $m \,\tbunch \gg 1$, the spin is again unable to develop a large vertical component. 

\paragraph{Resonance}
The decoherence due to $m\, \tbunch \gg 1$ may be removed by a resonance technique, that is by engineering $\omsm = m$. 
In this case, the rotation of the spin in the $xy$-plane occurs at the same frequency as the oscillation of $\omdm(t)$, and as a consequence $\vec{\omega}_\x{dm} \times \vec{S}$ does not change sign over the course of a single muon bunch. 
The spin will steadily rotate out of the $xy$-plane. 
Near-resonance, $\omsm \approx m$, the trajectory may be found by decomposing the harmonic perturbation $\vec{\omega}_\x{dm} = \omdm \cos(m t) \, \hat{x}$ into two counter-rotating perturbations, one clockwise and the other counter-clockwise in the $xy$-plane. 
One of the these circular components rotates with $\vec{S}$ and dominates the dynamics. 
Ignoring the other component and transforming to a frame rotating at $m$ yields a frame in which the precession frequency is constant and the spin trajectory may be easily found. 
Transforming back to the RMRF, the vertical component is 
\begin{align}
\label{eq:Sz-resonance}
 S_z \approx \frac{S \, \omdm \sin{\alpha}}{\sqrt{\omdm^2 + \lp m - \omsm \rp^2}} \,
  \sin \lp t \sqrt{\omdm^2 + \lp m - \omsm \rp^2} \rp.
\end{align}
For $m = \omsm$, this recovers a form similar to the static, spin frozen case of Eqn.~\eqref{eq:Sz-frozen-static}.
Again the vertical oscillation on-resonance is uniform over one bunch, however its amplitude will vary and may change sign between bunches. 
This is because at the start of a new bunch the spin is initialized to lie along $\hat{y}$, which differs from the position that a spin from the prior bunch would have if it survived until the start of the new bunch. 


\section{Sensitivity}
\label{sec:sensitivity}

In this section we determine the sensitivity of existing and upcoming muon precession experiments to the generic harmonic DM perturbations given in Section~\ref{sec:DM-precession}.
Such a DM signal may appear in muon precession data in three distinct ways:
\begin{enumerate}
   \item A time-resolved analysis of the ensemble of single-bunch signals may directly reveal temporal variation in the muon precession frequency $\voma(t)$. 
   \item Temporal variation of $\voma(t)$ may cause the stacked data to noticeably deviate from the expected harmonic behavior described in Section~\ref{sec:expected-signals}.
   \item The stacked data may follow the harmonic forms of Section~\ref{sec:expected-signals} within current precession, but the observed frequency or precession tilt may receive a measurable contribution which depends on the local DM density. 
 \end{enumerate} 
The first of these is the most compelling and provides an opportunity for DM detection upon reanalysis of existing and future muon precession data. 
The second and third allow us to set limits on DM-muon interactions using published, stacked results, while the third may also provide an explanation of the g-2 anomaly observed at BNL. 
A DM-muon interaction may give rise to one or more of these three signals, depending on the form of the interaction and the timescale of the perturbation, i.e.~the DM mass, relative to the various experimental timescales outlined in Section~\ref{sec:exp-deets}.

We begin with the signals and constraints resulting from the total electron count, which is applicable to g-2 experiments.
We then consider the vertical count, which applies to both g-2 and future frozen spin experiments, and which admits a resonant enhancement. 
Many of the derivations for the vertical count follow closely an analogous total count derivation, in which case only the final result is given. 
These results are applied to specific DM candidates in Section~\ref{sec:canidates}.

\subsection{Total Count}
\label{sec:sen-tc}
Ultralight DM may generate a frequency modulation or a frequency shift in the total count, as in Eqns.~\eqref{eq:Sp-parallel-harmonic} and~\eqref{eq:Sp-perp-wkb-harmoinc}.
We may describe both cases as a DM-induced frequency modulation of amplitude $\delta\omega$ and frequency $m$ in the oscillation of the momentum-component of spin $S_y$. 
A static frequency shift simply corresponds to $m=0$. 
During the $i^\x{th}$ muon bunch this has the form
\begin{align}
\label{eq:FM-bunch}
 S_{y,i}(t) &= S \cos\lp \omsm t +
  \frac{\delta\omega}{m} \big[ \sin(m t + \alpha_i) - \sin(\alpha_i) \big] \rp
\end{align} 
where $\alpha_i$ is the phase of the DM oscillation at the start of the $i^\x{th}$ bunch.
The stacked signal is 
\begin{align}
\label{eqn:stacked-Sy}
 \langle S_y \rangle = \frac{1}{N_b} \Sigma_i S_{y,i} 
\end{align}
where $N_b \approx 10^6 \,-\,10^8$ is the number of bunches observed per experimental run. 
Note that $\delta\omega$ is distinct from the DM contribution to the vector precession frequency $\vec{\omega}_\x{dm}$ and $m$ is distinct from the DM particle mass $\mdm$. 
$\delta\omega$ may scale either linearly or quadratically with the magnitude $|\vec{\omega}_\x{dm}|$, and $m$ may be equal to either $\mdm$, a non-zero multiple of $\mdm$, or it may vanish, depending on the form of the DM-muon interaction (see Section~\ref{sec:DM-precession}).

\subsubsection{Static Frequency Shift}
\label{sec:static-shift-limit}
A DM-induced shift in the precession frequency may be directly compared with the stacked results of muon precession experiments and the predicted SM value. 
The current discrepancy between theory and experiment makes this comparison more intriguing. 
The BNL experiment has measured $\omega_a$ with a precision $\sigma_{\omega_a} \approx 0.5 \cdot 10^{-6} \omega_a$ and found a discrepancy $\Delta \omega_a$ between their measurement and the SM prediction of $\Delta \omega_a = 3.3\,\sigma_{\omega_a}$~\cite{Bennett:2006fi}.
For a DM candidate which generates a frequency shift $\delta\omega$, we may immediately say the following:
\begin{enumerate}
 \item If $\delta\omega > \Delta \omega_a + \sigma_{\omega_a}$, this candidate is disfavored\footnote{Such a candidate is not properly excluded, as other new physics may provide an opposite and finely-tuned contribution to the precession frequency.} by at least 1-sigma. 
 \item If $\delta\omega < \sigma_{\omega_a}$, the candidate is unconstrained by this observable.
 \item If $\delta\omega$ lies within $\sigma_{\omega_a}$ of $\Delta \omega_a$, it provides a 1-sigma explanation of the discrepancy. 
 \item In the window $\sigma_{\omega_a} < \delta\omega < \Delta \omega_a - \sigma_{\omega_a}$, a candidate cannot be said to be disfavored nor would it explain the discrepancy.
 Such a candidate would provide a non-negligible contribution to $\omega_a$, but additional physics would be needed to fully explain the discrepancy. 
\end{enumerate}
These criteria are used for the constraints given in Section~\ref{sec:canidates}.
The Fermilab and J-PARC measurements anticipate a decrease in $\sigma_{\omega_a}$ by a factor of $4$ (see Section~\ref{sec:exp-deets}), and of course may yield a change in $\Delta \omega_a$, which will necessitate a slight update to those limits. 

\subsubsection{Stacked Envelope}
\label{sec:stackenv}
To what extent is a modulation with $m > 0$ visible in the stacked signal?
Averaging a collection of near-harmonic signals with similar frequencies will generically produce another near-harmonic signal whose frequency is an average of the individual frequencies and whose amplitude is given by an envelope that evolves at a rate given by the frequency spread of the individual signals. 
This is the phenomenon of beats. 
In our case, in the limit of a large number of bunches and $m \, \trun \gg 1$, the stacked signal $\langle S_y \rangle$ is given by the average of Eqn.~\eqref{eq:FM-bunch} over the DM phase $\alpha$. 
Here $\trun$ is the duration of a full experimental run, spanning all of the bunches in the stack. 
This average may be done exactly, yielding~\footnote{
The observed signal contains an additional exponential envelope due to muon decay, given in Eqn.~\eqref{eq:NE-general}. 
However, it is sufficient here to consider the average of the oscillatory factor $S_y$.}
\begin{align}
\label{eq:FM-average}
 \langle S_y \rangle &\approx \frac{S}{2\pi} \int_0^{2\pi} d\alpha \, 
  \cos\lp \omsm t + \frac{\delta\omega}{m} 
  \big[ \sin(m t + \alpha) - \sin(\alpha) \big] \rp \\
\label{eq:FM-stack}
  &= S \cos\lp \omsm t \rp 
  J_0 \lp 2\,\frac{\delta\omega}{m} \, 
  \left|\,\sin\lp\frac{m t}{2}\rp\right| \rp,
\end{align}
whereas the expected SM signal is $\langle S_y \rangle = S \cos\lp \omsm t \rp$.

The envelope in Eqn.~\eqref{eq:FM-stack} has the form of an additional decay of the signal. 
Such a decay would be noticed if sufficiently strong, however there is already present in the data a systematic effect which mimics this --- muons escaping the orbital trap.
These muon losses are found empirically at BNL to be $f_\x{loss}\approx 10\%$~\cite{Miller:2012opa}. 
We estimate that a stacked envelope will go unnoticed if it decays by no more than a fraction $f_\x{loss}$ over the span of the stacked bunch. 
This bounds the argument of the Bessel-envelope in Eqn.~\eqref{eq:FM-stack} to be $\lesssim f_\x{loss}$.
For simplicity, we implement this constraint as yielding an allowable DM candidate if
\begin{align}
\label{eq:envelope-condition}
 \delta\omega &\lesssim \frac{2 \, f_\x{loss}}{\tbunch} \,
 \x{Max}\lp \frac{1}{2} \, m\,\tbunch, \, 1 \rp
\end{align}
where $\tbunch$ is the bunch duration.
If the modulation does not vary appreciably over a bunch duration, this bounds the modulation amplitude in the g-2 experiments to be smaller than $\sim 10^{-3} \,\omsm$. 
For larger $m$ this weakens, as the envelope decay saturates due to the decoherence between the bunches.

\subsubsection{Stacked Frequency Residual}
\label{sec:stackfreqres}
Supposing that Eqn.~\eqref{eq:envelope-condition} is satisfied, the stacked signal $\langle S_y \rangle$ takes the form of a harmonic oscillation. 
The frequency of this oscillation is approximately $\omsm$, but only in so far as the discrete average of the bunches approximates the continuous, single-period average over DM phase of Eqn.~\eqref{eq:FM-average}. 
Given Eqn.~\eqref{eq:envelope-condition}, the discrete average is well-approximated by
\begin{align}
\label{eq:FM-discrete-average}
 \langle S_y \rangle &\approx S \, \cos\lp \omsm t + 
  \frac{\delta\omega}{m} \, \frac{1}{N_b} \Sigma_i 
  \big[ \sin(m t + \alpha_i) - \sin(\alpha_i) \big] \rp .
\end{align}
This follows from linearizing Eqn.~\eqref{eq:FM-bunch} in the DM-induced phase shift.

We will be primarily concerned with the case $m \tbunch \ll 1$, where the modulation is approximately static over a single bunch. 
Then we have, 
\begin{align}
\label{eq:FM-discrete-average-staticbunch}
 \langle S_y \rangle &\approx S \, 
 \cos\lp \lb \omsm + 
  \delta\omega \, \frac{1}{N_b} \Sigma_i \cos\lp \alpha_i \rp \rb t \rp ,
\end{align}
that is, the stacked frequency is simply the average of the frequencies of each bunch. 
Note that $\alpha_i = \alpha_0 + m t_i$, where $t_i$ is the starting time of the $i^\x{th}$ bunch. 
In most of our regime of interest, the average time between bunches $t_{i+1} - t_i$ is short compared to the modulation period $m^{-1}$, so the discrete average in Eqn.~\eqref{eq:FM-discrete-average-staticbunch} may be approximated by an integral
\begin{align}
\label{eq:freq-residual-integral-phase}
  \frac{\delta\omega}{N_b} \Sigma_i \cos\lp \alpha_i \rp
  \approx \frac{\delta\omega}{\trun} \int_0^{\trun} dt \, \cos\lp m t + \alpha_0 \rp
  \sim \frac{\delta\omega}{\x{Max}\lp m \, \trun, 1 \rp},
  \;\;\;\;\; \lb \lp t_{i+1} - t_i \rp m \ll 1 \rb. 
\end{align} 
where $\trun$ is the duration of the entire data-taking run, encompassing all bunches.
If $\lp t_{i+1} - t_i \rp m \gtrsim 1$, the value of the discrete average of frequencies depends on the uniformity of the time interval between bunches.
We assume that the duration of this interval may vary by $\OO \lp 1 \rp$ between different pairs of bunches, in which case the discrete average becomes well-approximated by a random-walk,
\begin{align}
\label{eq:freq-residual-integral-rw}
  \frac{\delta\omega}{N_b} \Sigma_i \cos\lp \alpha_i \rp
  \approx \frac{\delta\omega}{\sqrt{N_b}},
  \;\;\;\;\; \lb \lp t_{i+1} - t_i \rp m \gtrsim 1 \rb .
\end{align}
Taking the time interval between bunches to be given on average by $\trun/N_b$, the full result is 
\begin{align}
\label{eq:freq-residual-integral}
  \frac{\delta\omega}{N_b} \Sigma_i \cos\lp \alpha_i \rp
  \approx \frac{\delta\omega}{\x{Min}\lb\x{Max}\lp m \, \trun, 1 \rp, \sqrt{N_b}\rb} . 
\end{align}

This stacked frequency shift coincides with the static $m=0$ case if $m \trun \ll 1$, for which the shift is simply $\sim \delta\omega$ as in Section~\ref{sec:static-shift-limit}.
For larger $m$ this is suppressed as the DM oscillation averages out. 
The suppressed shift is still constrained in the same manner as described in Section~\ref{sec:static-shift-limit}.
A DM candidate is allowed if 
\begin{align}
\label{eq:residual-limit}
 \delta\omega &\lesssim 4 \, \sigma_{\omega_a} \,
 \x{Min}\lb\x{Max}\lp m \, \trun, 1 \rp, \sqrt{N_b}\rb.
\end{align}
Note that frequency residual limit in Eqn.~\eqref{eq:residual-limit} is generally less constraining than the envelope limit considered above in Eqn.~\eqref{eq:envelope-condition}, as the DM averaging effects appear at a much smaller value of $m$ for the frequency residual than they do for the envelope decay. 
Only for $m \lesssim 10^{-20}~\eV$ does the frequency residual give the stronger limit.

\subsubsection{Time-Resolved Frequency Tracking}
\label{sec:time-resolved-freq-sensitivity}
A DM modulation with $m > 0$ may be directly revealed by a time-resolved analysis of muon precession using each unstacked bunch. 
There are many specific analysis techniques that one might use, and it is beyond the scope of this work to assess them in detail. 
We are concerned instead with understanding the general sensitivity of the g-2 data to a DM modulation signal. 
For simplicity we focus on the case $m \, \tbunch \lesssim 1$, corresponding to $m \lesssim 10^{-12}~\eV$ for the BNL and Fermilab experiments, for which the modulated precession frequency is constant over the duration of one bunch. 
The opposite limit, $m \, \tbunch \lesssim 1$, may be probed as well with an analysis of modulation occurring within each bunch, however we leave that case to future work. 

For $m \, \tbunch \lesssim 1$, one may determine a local precession frequency $\omega(t_i)$ for each bunch, where $t_i$ is the start time of the $i^\x{th}$ bunch. 
This may be done by fitting independently the oscillations observed in each bunch.
The modulated precession frequencies $\omega(t_i)$ depend on the DM field, so this is a direct measurement of a possible DM background interacting with muons. 
Consider the Fourier spectrum $\tilde{\omega}(\Omega)$ of the time series $\omega(t_i)$. 
We denote the frequency of this spectrum as $\Omega$, to avoid confusion with the precession frequency itself $\omega(t_i)$. 
The zero-mode of this spectrum is non-vanishing and corresponds to $\omsm$. 
We may normalize $\tilde{\omega}$ as 
\begin{align}
 \label{eq:FT-of-frequency}
 \tilde{\omega}\lp\Omega\rp = \frac{1}{N_b} \, \Sigma_i \, \omega(t_i) \, e^{- i \Omega t_i}
\end{align}
so the zero-mode is indeed $\tilde{\omega}(0) \approx \omsm$. 
A DM-induced modulation of the form of Eqn.~\eqref{eq:FM-bunch} appears in the spectrum as a peak of height $\delta\omega$ at $\Omega = m$. 

This DM signal is detectable provided $\delta \omega$ is sufficiently large relative to the noise in $\tilde{\omega}$.
The fit which determines $\omega(t_i)$ differs from the fit done on the stacked data, described in see Section~\ref{sec:exp-data-analysis}, only in the number of counts and thus the SNR of the individual bunch. 
The precision of such a fit scales inversely with the square root of the number of counts~\cite{Miller:2012opa}, so the noise in $\omega(t_i)$ is white and has an amplitude $\sigma_i \sim \sigma_{\omega_a} N_b^{1/2}$, where $\sigma_{\omega_a}$ is the precision of the fit to the stacked signal and $N_b$ is the number of bunches. 
For the Fermilab and J-PARC measurement, $\sigma_{\omega_a} \approx 10^{-7} \, \omsm$ and $\sigma_i \sim 10^{-3} \, \omsm$. 
The noise in each frequency bin of $\tilde{\omega}$ is thus $\sigma_{\omega_a}$.
This is sensible, as the stacked analysis corresponds to measuring the height of the peak in the spectrum at $\Omega = 0$. 
The remaining modes $\Omega > 0$ are currently unused, but may be utilized for a DM search. 

The specific frequency modes $\Omega_i$ to which g-2 data is sensitive is determined by the specific timing intervals of the bunches. 
This is complicated by the fact that the bunches are not uniformly spaced in time, and a full analysis requires knowledge of the intervals between each bunch.
This is beyond the scope of the present work. 
We seek an estimate of the sensitivity of such an analysis, and for our purposes we simply take the bunches to be uniformly spaced by their average spacing, $\trun/N_b$. 
Then $\tilde{\omega}(\Omega)$ probes modes spaced by $\trun^{-1}$ with a maximum frequency of $N_b\,\trun^{-1}$. 
These correspond to DM masses of $10^{-23}~\eV$ and $10^{-15}~\eV$, respectively. 
The approximation of a uniform interval between bunches has little effect on $\tilde{\omega}(\Omega)$ at small $\Omega$, but it sets the value of the maximal frequency $N_b\,\trun^{-1}$. 
In a full analysis, sensitivity will extend beyond $N_b\,\trun^{-1}$ as some bunches are spaced much closer together than the average spacing. 

The detection reach may then be estimated as follows. 
The DM modulation peak has a width $\delta \Omega \approx m v_\x{dm}^2 \approx 10^{-6} \, m$, due to the finite width of the DM velocity distribution. 
If $m v_\x{dm}^2 < \trun^{-1}$ then the DM oscillation is coherent over the course of an experimental run, or equivalently the DM peak in $\tilde{\omega}$ lies entirely within a single frequency bin. 
The SNR of that bin is $\x{SNR} = \delta\omega/\sigma_{\omega_a}$. 
If $m v_\x{dm}^2 > \trun^{-1}$ then the phase of the DM oscillation will drift during the course of a run, and the resulting peak in the spectrum will span several frequency bins. 
The full SNR is now properly given by the quadrature-sum of the SNR of each of those bins, which is $\x{SNR} = \lp m v_\x{dm}^2 \trun \rp^{-1/2} \, \delta\omega/\sigma_{\omega_a}$.
The SNR covering both regimes is 
\begin{align}
 \x{SNR} = \frac{\delta\omega}{\sigma_{\omega_a}} \, 
 \frac{1}{\x{Max}\lp m v_\x{dm}^2 \trun, 1 \rp^{1/2}}.
\end{align}
We take the detection reach to be given by $\x{SNR} > 3$. 
This is properly the reach only for a predetermined frequency $m$, which is of interest in the event that a candidate DM signal is found in other experiments. 
Accounting for the look-elsewhere effect in a search with no preferred modulation frequency requires taking $\x{SNR} \gtrsim 15$, with the exact threshold depending on the desired confidence. 
This amounts to a sensitivity which is about a factor of $\sim 5$ worse than those shown in Section~\ref{sec:canidates}. 

\subsection{Vertical Count}
\label{sec:sen-vc}
A non-zero vertical count is generated only for perpendicular frequency perturbations.
We consider here a harmonic DM signal of frequency $m$ in the non-resonant case, which in the $i^\x{th}$ muon bunch is given by (see Section~\ref{sec:perpendicular-perturbation})
\begin{align}
 \label{eq:FM-perp-bunch}
 S_{z,i} &\approx S \frac{\omdm}{\omsm} \cos\lp m t + \alpha_i \rp
 \sin\lp \bar{\omega} t + \Phi_i\lb t \rb \rp \\ 
 \x{where:}\;\; 
  \bar{\omega} &= \omsm + \frac{1}{4} \frac{\omdm^2}{\omsm} \\
  \Phi_i \lb t \rb &= 
    \frac{1}{8} \frac{\omdm^2}{\omsm m} \lb 
    \vphantom{\frac{\omdm^2}{\omsm m}} \sin(2 m t + 2 \alpha_i) - \sin(2 \alpha_i) \rb 
\end{align}
where $\alpha_i$ is the phase of the DM oscillation at the start of the $i^\x{th}$ bunch and the stacked signal is 
\begin{align}
\label{eqn:stacked-Sz}
 \langle S_z \rangle = \frac{1}{N_b} \, \Sigma_i \, S_{z, i}. 
\end{align}
The limits and detection reach in this case are analogous to those for the total count in Section~\ref{sec:sen-tc}, with the distinction that in this case it is the amplitude, not the frequency, of the precession which is observed and the DM oscillation induces an amplitude modulation in the signal rather than a frequency modulation. 
In addition, as demonstrated in Section~\ref{sec:DM-precession}, this signal is always accompanied by a static frequency shift in the total count of amplitude $\delta\omega = \omdm^2/8\omsm$, which is subject to the constraints of Section~\ref{sec:sen-tc}.
That is, 
\begin{align}
\label{eq:perp-DC-constraint}
 \omdm \lesssim \lp 8 \, \sigma_{\omega_a} \, \omsm \,
   \x{Min}\lb\x{Max}\lp m \, \trun, 1 \rp, \sqrt{N_b}\rb \rp^{1/2}.
\end{align}
At its most stringent, this is $\omdm \lesssim 3\cdot 10^{-3}\,\omsm$ for the g-2 experiments.

\subsubsection{Stacked Amplitude Residual}
\label{sec:stackampres2}
For a perpendicular perturbation which satisfies Eqn.~\eqref{eq:perp-DC-constraint}, the stacked vertical signal $\langle S_z \rangle$ is well approximated by 
\begin{align}
 \langle S_z \rangle \approx 
 \frac{1}{N_b} \Sigma_i \, \cos\lp m t + \alpha_i \rp \cdot
 S \, \frac{\omdm}{\omsm}\, \sin\lp \bar{\omega} t \rp. 
\end{align}
We have ignored the frequency modulation, as in this case it is subdominant to the amplitude modulation.
The stacked amplitude is given by an average over samples of a sinusoid, analogous to the frequency residual in Eqn.~\eqref{eq:freq-residual-integral}.   
The typical stacked signal is thus
\begin{align}
 \langle S_z \rangle \approx 
 S \, \frac{\omdm}{\omsm} \frac{1}{\x{Max}\lp m \, \trun, 1 \rp} \cdot 
 \, \sin\lp \bar{\omega} t \rp. 
\end{align}
Let $\sigma_{\perp}$ be the sensitivity of a static EDM search to the perpendicular component of precession frequency.
For the existing BNL measurement, $\sigma_\perp \approx 0.5 \cdot 10^{-3} \omsm$ (see Section~\ref{sec:exp-deets}). 
The sensitivity to the amplitude of a vertical oscillation is $\sigma_\perp S /\omsm$ and the null result of BNL implies that a DM candidate is allowed only if 
\begin{align}
 \omdm \lesssim \sigma_\perp \, \x{Max}\lp m \, \trun, 1 \rp
\end{align}

\subsubsection{Time-Resolved Amplitude Tracking}
\label{sec:timeresamp2}
It is again possible to use a time-resolved analysis of the unstacked bunches to reveal the modulation induced by a DM background. 
As in Section~\ref{sec:time-resolved-freq-sensitivity}, we consider here the general sensitivity in the limit that $m \, \tbunch \lesssim 1$, where the precession is approximately uniform for the duration of each bunch. 

We employ the same strategy outlined in Section~\ref{sec:time-resolved-freq-sensitivity}, fitting each bunch independently and then considering the Fourier spectrum of the outcome of those fits. 
In this case, the signal is expected to be of the form of Eqn.~\eqref{eq:FM-perp-bunch} in each bunch and the quantity of interest is the amplitude modulation. 
We may fit each bunch to the form
\begin{align}
\label{eq:Sz-fitting-func}
  S_{z}^\x{fit} = A \, S \, \sin\lp \bar{\omega} t + \phi \rp
\end{align}
for the amplitude $A$ and construct a time series $A(t_i)$, where $t_i$ is the start time of the $i^\x{th}$ bunch. 
The total count will oscillate at the same frequency $\bar{\omega}$ and with a fixed phase shift relative to the vertical count (see Eqns.~\eqref{eq:Sp-perp-wkb-harmoinc} and~\eqref{eq:SB-perp-wkb-harmoinc}). 
Thus the frequency and phase in Eqn.~\eqref{eq:Sz-fitting-func} may be determined by first fitting the higher-SNR total count, and the vertical count can be fit for only the amplitude $A$.
Note that this is again the same procedure currently applied to the stacked signal, as described in Section~\ref{sec:exp-data-analysis}, but now applied independently to each bunch. 

We may consider the Fourier spectrum $\tilde{A}(\Omega)$ of $A(t_i)$, normalized as: 
\begin{align}
 \label{eq:FT-of-vertical-amplitude}
 \tilde{A}\lp\Omega\rp = \frac{1}{N_b} \, 
 \Sigma_i \, A(t_i) \, e^{- i \Omega t_i}.
\end{align}
The DM modulation now appears as a peak of height $\omdm/\omsm$ at frequency $\Omega = m$. 
By an analogous argument to that given in Section~\ref{sec:time-resolved-freq-sensitivity}, the noise amplitude in each frequency bin of $\tilde{A}(\Omega)$ is $\sigma_\perp/\omsm$ and the SNR of a DM modulation is 
\begin{align}
\label{eq:tr-vertical-snr}
 \x{SNR} = \frac{\omdm}{\sigma_\perp} \, 
 \frac{1}{\x{Max}\lp m v_\x{dm}^2 \trun, 1 \rp^{1/2}}.
\end{align}
For the upcoming Fermilab and J-PARC experiments, $\sigma_\perp \approx 0.5 \cdot 10^{-5} \omsm$. 
We set the threshold SNR for detection as in Section~\ref{sec:time-resolved-freq-sensitivity}.

\subsubsection{Frozen Spin}
\label{sec:sen-frozen}
For a frozen spin experiment, we consider an analogous time-resolved measurement to that of Section~\ref{sec:timeresamp2}.
In the limit $m \, \tbunch \ll 1$, the signal has the form of Eqn.~\eqref{eq:Sz-frozen-static}. 
$\omdm$ is generally small, so that this is a signal which grows linearly in time, 
\begin{align}
\label{eq:Sz-frozen-linear}
 S_z(t) \approx S \omdm \cos \lp \alpha \rp \, t .
\end{align}
Simply averaging $S_z$ over each bunch yields a signal $\bar{S}_z(t_i)$ which oscillates between bunches according to the DM phase $\alpha$, 
\begin{align}
 \bar{S}_z(t_i) \approx \frac{S}{2} \omdm \tbunch \, \cos \lp \alpha_i \rp .
\end{align}
As in Section~\ref{sec:stackampres2}, let $\sigma_\perp$ be the sensitivity of a spin frozen experiment to a static, perpendicular precession frequency. 
From the Fourier spectrum of $\bar{S_z}(t_i)/S$, the SNR of a DM modulation peak of frequency $m$ is 
\begin{align}
\label{eq:SNR-frozen-nearstatic}
 \x{SNR} =  \frac{\omdm}{\sigma_\perp} \, 
 \frac{1}{\x{Max}\lp m v_\x{dm}^2 \trun, 1 \rp^{1/2}},
 \;\;\;\; \lb m \, \tbunch \ll 1\rb,
\end{align}
which follows from an analogous argument to that of Sections~\ref{sec:time-resolved-freq-sensitivity} and~\ref{sec:timeresamp2}.
For larger masses, $m \, \tbunch \gg 1$, the signal follows Eqn.~\eqref{eq:Sz-frozen-nonstatic} and the average over one bunch is suppressed:
\begin{align}
 \bar{S}_z(t_i) \approx - S \frac{\omdm}{m} \sin \lp \alpha_i \rp ,
 \;\;\;\; \lb m \, \tbunch \gg 1\rb.
\end{align}
The SNR covering both regimes is 
\begin{align}
\label{eq:SNR-frozen-full}
 \x{SNR} = \frac{\omdm}{\sigma_\perp} \,
 \frac{1}{\x{Max}\lp m v_\x{dm}^2 \trun, 1 \rp^{1/2} \,\, 
 \x{Max}\lp m \tbunch, 1 \rp}
\end{align}
and we set the threshold SNR for detection as in Section~\ref{sec:time-resolved-freq-sensitivity}.

\subsubsection{Resonance}
\label{sec:sen-res}
The amplitude of the vertical signal is enhanced if the DM modulation frequency $m$ matches the SM rotation of the spin $\omsm$. 
For an experiment operating with fixed external fields and muon momentum, this results in an extended detection reach for perpendicular perturbations in a narrow frequency window around $m = \omsm$. 
In the previous and upcoming g-2 experiments, this corresponds to $m \approx 10^{-10} \, \x{eV}$. 
Following Eqn.~\eqref{eq:Sz-resonance}, on resonance, the vertical spin component will grow linearly during each bunch, as the bunch duration is short compared to the on-resonance precession frequency of the spin. 
The angular spin velocity will vary between bunches according the DM phase, analogous to the frozen spin signal given in Eqn.~\eqref{eq:Sz-frozen-linear}. 
Following the time-resolved analysis procedure of Section~\ref{sec:sen-frozen}, the near-resonance SNR of this signal is 
\begin{align}
\label{eq:tr-resonace-snr}
 \x{SNR} = \frac{\omsm \tbunch}{2} \, \frac{\omdm}{\sigma_\perp} \, 
 \frac{1}{\x{Max}\lp m v_\x{dm}^2 \trun, 1 \rp^{1/2}}.
\end{align}
This SNR is enhanced by a factor $\omsm \tbunch \approx 100$ relative to the non-resonant SNR of Eqn.~\eqref{eq:tr-vertical-snr}. 
The reach is thus extended to $\omdm/\omsm \gtrsim 10^{-8}$ for the upcoming Fermilab and J-PARC measurements.  
From Eqn.~\eqref{eq:Sz-resonance}, the frequency width of this enhancement is given by $| m - \omsm | < 1/ \tbunch \approx 10^{-2}\,\omsm$. 
This is very narrow compared to the range of $\mdm$ considered in Section~\ref{sec:canidates}, and so we refrain from showing this peak in sensitivity in Figs.~\ref{fig3}, ~\ref{fig4}, ~\ref{fig5}, and~\ref{fig6}. 

In addition to yielding a fixed sensitivity peak near $m = \omsm$ in spin precession experiments, resonance may be used to extend the reach of a future DM search at a variety of frequencies by tuning $\omsm$ to a desired search window. 
This would be useful for follow-up observations in the event that an ultralight DM signal is observed in other experiments. 
The most natural and sensitive setup for such a search is the proposed frozen spin EDM experiments, which plan to employ electric fields to tune $\omsm$ and utilize future high-intensity muon sources (see Section~\ref{sec:exp-deets}).
Then sensitivity of such a search matches that of a near-static frozen spin signal, given in Eqn.~\eqref{eq:SNR-frozen-nearstatic}, as the resonant signal follows the same form as the non-resonant static signal. 
We show this reach in Section~\ref{sec:canidates} for all $\mdm$, indicating the peak reach of a narrow resonant search at the given $\mdm$. 
In principle a future search may cover a wide range of $\mdm$ by systematically varying $\omsm$, in which case the sensitivity is as shown in Section~\ref{sec:canidates}. 
There are important practical challenges to varying $\omsm$ over a large range, which are beyond the scope of this work. 
The results of Section~\ref{sec:canidates} represent the ideal limit of such an experiment.


\section{Candidates}
\label{sec:canidates}

In this section, we explore models of ultralight dark matter that would produce one of the signals enumerated in Sections~\ref{sec:DM-precession} and~\ref{sec:sensitivity}. We consider models where the ultralight boson couples preferentially to muons so as to avoid strong tension with experiments and limits on couplings to electrons, photons, and nucleons. In the absence of a symmetry, the muon coupling will radiatively generate couplings to other SM particles. In this Section, we conservatively  project only direct muon constraints and postpone a discussion of indirect constraints from radiatively generated couplings and fine-tuning, which are severe for models without a shift symmetry or gauge symmetry, to Appendix.~\ref{looplevel}. 

\subsection{Scalars}

\subsubsection{$\phi \bar{\mu} \mu$} 
\label{sec:can-yukawa}

The scalar coupling we first consider is 
\begin{equation}
\mathcal{L}\supset y \, \phi \bar{\mu} \mu
\end{equation}
This operator has already been proposed to explain the muon g-2 anomaly (see for e.g. \cite{Chen:2017awl} and references therein), albeit through radiative corrections to muon g-2. This limits $y \lesssim 10^{-3} $ for small enough $m_\phi$. Constraints could also be drawn from the anomalous cooling of SN1987A \cite{Brust:2013xpv,DEramo:2018vss} owing to the presence of a non-trivial amount of muons inside. Finally, it may also result in 5th force constraints from neutron stars \cite{Dror:2019uea}. These, however, suffer from uncertainties in the muon abundance inside the neutron star and moreover can be avoided by introducing a quadratic coupling to nuclei, $\phi^2 \bar{n}{n}$, which effectively screens the fifth force. There are also indirect constraints from couplings introduced at loop level which we discuss in Appendix.~\ref{looplevel}. 

If this scalar $\phi$ is DM, it induces an oscillating mass for the muon
\begin{equation}
m_\mu = m_\mu^{\rm SM} + y \sqrt{\frac{2\rho_{\phi}}{m_{\phi}}}
  \cos\lp m_\phi t\rp
\end{equation}
$\omega_{\rm sm}$ depends on $m_\mu$ through Eqn.~\ref{eq:omega_a}. Expanding in small $y$, we get,
\begin{equation}
\vec{\omega}_{\rm dm}=\vec{\omega}_{\rm sm} \frac{y}{m_\mu} 
  \sqrt{\frac{2\rho_{\phi}}{m_{\phi}}} \cos \lp m_\phi t \rp
\end{equation}
This is a parallel perturbation as discussed in Sec.~\ref{sec:parallel-perturbation}. 

\begin{figure}[htpb]
\centering
\includegraphics{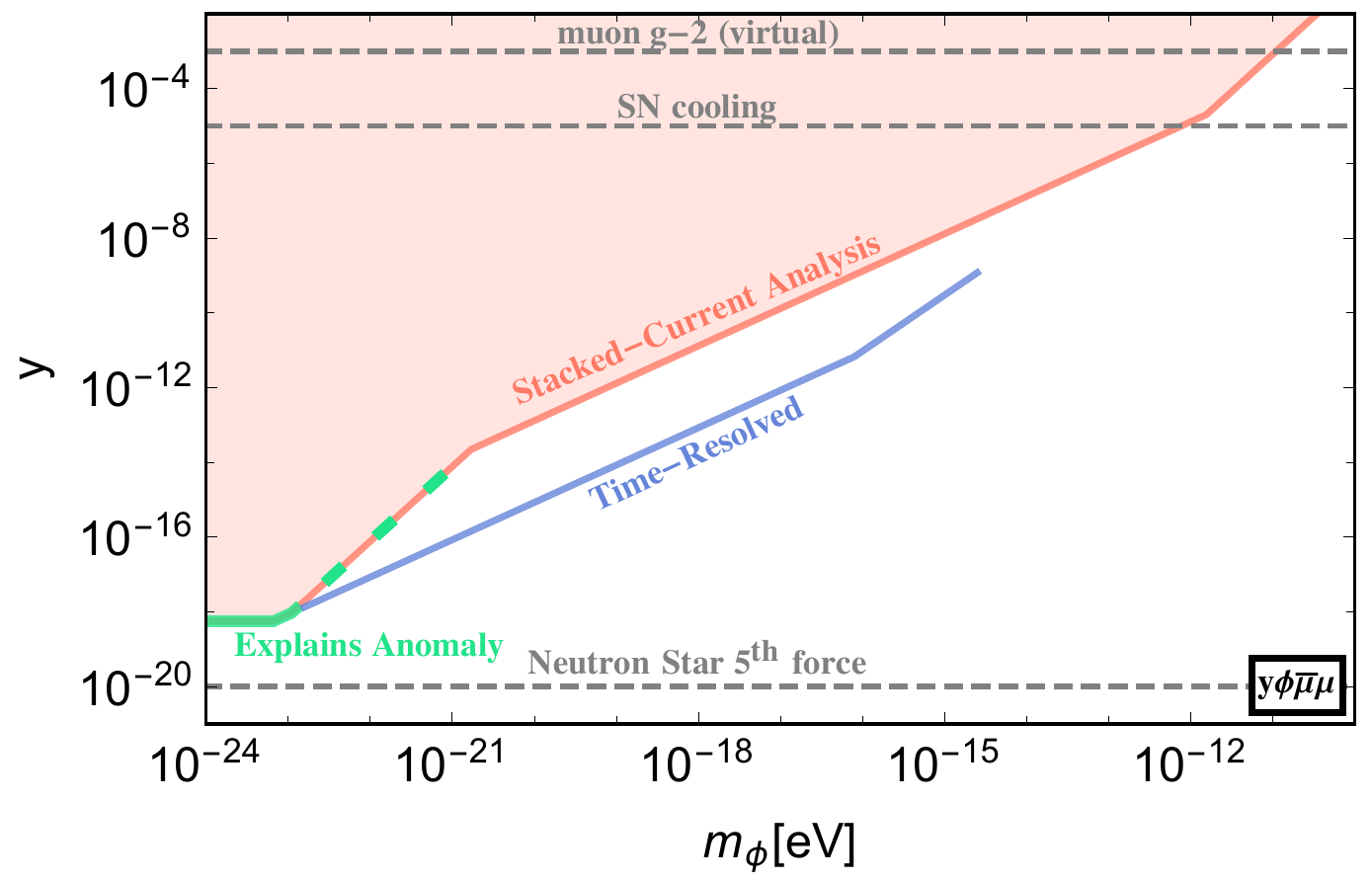}
\caption{Limits and projections for a scalar DM candidate $\phi$ with Yukawa coupling $y \phi \bar{\mu}{\mu}$, from current and future muon precession experiments are displayed. The red shaded region corresponds to deviations to the stacked analysis that would have already been seen in the g-2 analysis. The green (dashed) line corresponds to parameter space that can explain the observed g-2 anomaly with ($12.5\%$) $50\%$ probability. Shown in blue are projections for a time-resolved analysis. Shown in gray are constraints from virtual corrections to muon g-2 \cite{Chen:2017awl}, SN cooling adapted from \cite{Bollig:2020xdr} and 5th force constraints from NS \cite{Dror:2019uea}. See Section~\ref{sec:can-yukawa} for details.}
\label{fig1}
\end{figure}

Constraints and projections for this operator from different experiments are plotted in Fig.~\ref{fig1}. The red shaded region corresponds to parameters that predict deviations not observed in the completed analysis at BNL and is ruled out at the $2\sigma$ level. At the smallest masses, the frequency shift is static as discussed in Sec.~\ref{sec:static-shift-limit}. However, the limit is flat as it is only the change in the effective mass of the muon between the muonium experiments and the g-2 experiment which is observable here. The boundary of this region marked in green could explain the anomaly with $50\%$ probability --- it happens in the event that the scalar vev decreases in magnitude from the muonium measurement to the g-2 measurement, resulting in a lower muon mass. At scalar masses corresponding to frequencies larger than $1\, \textrm{year}^{-1}$, the red shaded region corresponds to deviations in muon g-2 larger in magnitude but in principle different in sign over the three different BNL runs. For this reason, the boundary is green-hatched to indicate the low probability that the three runs reported the same sign deviation. At masses larger than $\sim 10^{-21}\,\x{eV}$, there is noticeable change to the decay envelope (Sec.~\ref{sec:stackenv}). At even higher masses, coherence is lost over a bunch and only stacked frequency residuals set a limit (Sec.~\ref{sec:stackfreqres}). If time stamps of individual electron events are retained and used for a time-resolved analysis as described in detail in Sec.~\ref{sec:time-resolved-freq-sensitivity}, a projected detection reach shown by the blue line is obtained. Also shown are constraints from the virtual contribution to the g-2 measurement, cooling from SN, and 5th force constraints from NS mergers in gray. 

\subsubsection{$\phi^2 \bar{\mu} \mu$} 
\begin{figure}[htpb]
\centering
\includegraphics{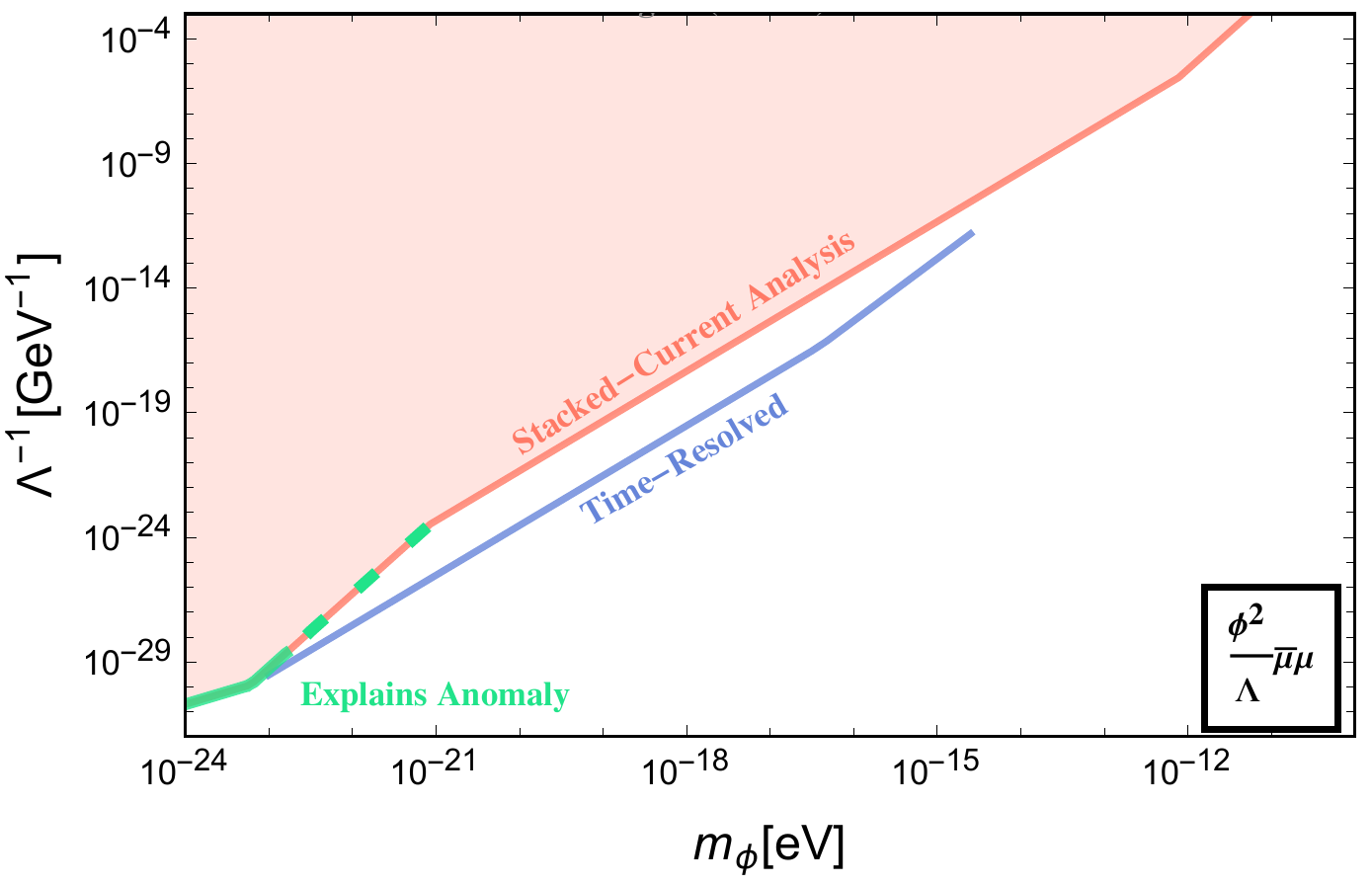}
\caption{Limits and projections for a scalar DM candidate $\phi$ with coupling $\frac{\phi^2}{\Lambda} \bar{\mu}{\mu}$ using the same color coding discussed in Fig.~\ref{fig1}.}
\label{fig2}
\end{figure}

In models where $\phi$ originally satisfies a $Z_2$ symmetry, we start with a Lagrangian, 
\begin{equation}
\mathcal{L}\supset \frac{1}{\Lambda} \phi^2 \bar{\mu} \mu
\end{equation}
This operator is not as well constrained as the Yukawa case as the scalar appears with additional loops or in pairs and hence its effect is suppressed. 
Repeating the analysis above, we obtain, 

\begin{equation}
\label{eq:omdm-phisq}
\vec{\omega}_{\rm dm}=\vec{\omega}_{\rm sm} \frac{2\rho_{\phi}\cos^2 \lp m_\phi t \rp}{m_{\phi} \Lambda m_\mu}=\vec{\omega}_{\rm sm} \frac{\rho_{\phi} }{m_{\phi}\Lambda m_\mu}\left(1+\cos \lb 2m_\phi t \rb \right)
\end{equation}
The constraints on this parameter space are derived similarly to the linear case and plotted in Fig.~\ref{fig2}.
Note that the constant term in Eqn.~\eqref{eq:omdm-phisq} does not contribute to the limits, as it is perfectly degenerate with the ``intrinsic" muon mass $m_\mu$.


\subsection{Pseudoscalars}

\subsubsection{$\partial_\alpha a \, \bar{\mu} \gamma^\alpha \gamma_5 \mu$}
\label{sec:can-linear-wind}

We start with the axion-muon ``wind" coupling,
\begin{equation}
\mathcal{L} \supset \frac{\partial_{\alpha} a}{\Lambda} \bar{\mu} \gamma^\alpha \gamma_5 \mu.
\end{equation}
In a background axion field $a$, this interaction generates a spin torque described in the muon rest frame by the Hamiltonian term~\cite{Graham:2013gfa}
\begin{align}
\label{eq:wind-hamiltionian}
  H \supset \frac{1}{\Lambda} \vec{\nabla} a \cdot \vec{S},
\end{align}
where $\vec{S}$ is the muon spin, and contributes an amount 
\begin{align}
\label{eq:wind-omega-rest}
  \vec{\omega}_r = \frac{1}{\Lambda} \; \vec{\nabla} a
\end{align}
to the muon's rest-frame precession frequency. 
In its rest frame the muon spin precesses about the direction of the axion momentum $\vec{p}_a$, as $\vec{\nabla} a \sim a \, \vec{p}_a$ for a plane wave axion mode. 

In Eqn.~\eqref{eq:wind-omega-rest}, $a$ is the axion field in the muon rest frame and the gradient is taken with respect to the rest frame coordinates. 
In the lab frame, the axion DM background is non-relativistic and has the form $a \approx a_0 \cos\lp m_a t \rp$ while the muon is relativistic. 
Thus in the muon rest frame the axion background is now relativistic and has momentum $\vec{p}_a \approx \gamma m_a \vec{v}$, where $\vec{v}$ and $\gamma$ are the velocity and boost factor respectively of the muon in the lab frame. 
Then $a \approx a_0 \cos\lp E \, t' - \vec{p}_a \cdot \vec{x}' \rp$ in the muon rest frame, and 
\begin{align}
 \vec{\omega}_r \approx -\frac{a_0}{\Lambda} \gamma m_a \vec{v} 
 \sin\lp E \, t' - \vec{p}_a \cdot \vec{x}' \rp
  = -\frac{a_0}{\Lambda} \gamma m_a \vec{v} 
 \sin\lp \mdm t \rp
\end{align}
where primes refer to muon rest frame coordinates and $t$ is the lab frame time. 
This gives a perpendicular frequency perturbation via Eqn.~\eqref{eq:most-generl-omegea_a},
\begin{align}
\label{eq:omdm-wind}
  \vec{\omega}_\x{dm} \approx 
  - \frac{a_0}{\Lambda} m_a \vec{v} \sin\lp \mdm t \rp
  = - \frac{\sqrt{2 \rho_\x{dm}}}{\Lambda} \, \vec{v} \sin\lp \mdm t \rp.
\end{align}
This perturbation is perfectly perpendicular as we have ignored the velocity of the axion DM in the lab frame. 
There is, in fact, also a parallel perturbation due to the DM velocity component along the vertical direction, however this is suppressed relative to Eqn.~\eqref{eq:omdm-wind} by at least $v_\x{dm} \approx 10^{-3}$ and we may ignore it. 

Direct constraints on this coupling come from virtual corrections to the measured muon g-2 (this produces a wrong-sign contribution to muon g-2 and hence does not explain the anomaly), which gives $\Lambda \ge 1$ TeV for small enough $m_a$~\cite{DEramo:2018vss}.
Constraints could also be drawn from the anomalous cooling of SN1987A \cite{Brust:2013xpv,DEramo:2018vss,Bollig:2020xdr} owing to the presence of a non-trivial amount of muons inside, yielding $\Lambda \ge 10^6$ GeV. However there are sizable uncertainties in the muon abundance inside supernovae which translate to large uncertainties in these limits. 

\begin{figure}[htpb]
\centering
\includegraphics{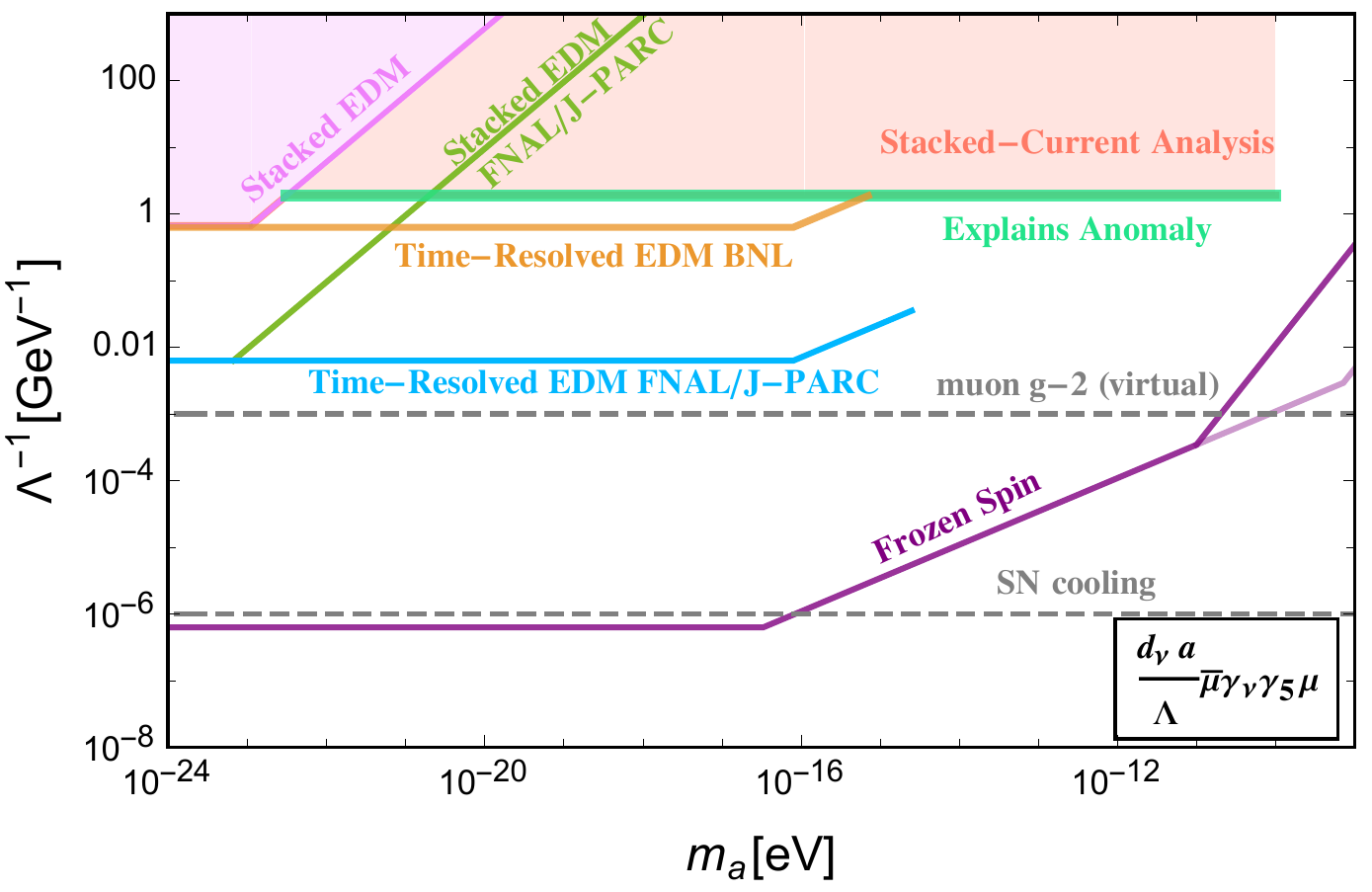}
\caption{Limits and projections for a pseudoscalar DM candidate $a$ with the wind coupling $ \frac{\partial_{\alpha} a}{\Lambda} \bar{\mu} \gamma^\alpha \gamma_5 \mu$, from current and future muon precession experiments are displayed. The red (pink) shaded region corresponds to deviations to the stacked analysis that would have already been seen in the BNL g-2 (EDM) analysis. The light green line corresponds to parameter space that can explain the observed g-2 anomaly. Shown in dark green, orange and blue are projections for stacked and time-resolved analyses of EDM at BNL and Fermilab/J-PARC. Frozen spin experiments have a projected detection reach shown in dark (light) purple for a static (resonant) measurement. Shown in gray are constraints from virtual corrections to g-2 \cite{DEramo:2018vss} and SN cooling \cite{Bollig:2020xdr} which effectively rule out a DM explanation to the g-2 anomaly from this operator}
\label{fig3}
\end{figure}

Constraints and projections for this operator are plotted in Fig.~\ref{fig3}. As explained in Sec.~\ref{sec:perpendicular-perturbation}, perpendicular perturbations are always accompanied by a static shift in the precession frequency which is positive definite. The green line corresponds to the parameter space that explains the anomaly and the region above marked in red would predict even larger $g-2$ measurements which are disfavored. The perpendicular perturbations can also be seen in the vertical count, and the non-observation of a static EDM rules out the pink region (see Sec.~\ref{sec:stackampres2} for more detail). If a time-resolved analysis is carried out, as outlined in Sec.~\ref{sec:timeresamp2}, the BNL and Fermilab/J-PARC data could be used to constrain regions above the orange and blue lines respectively. Finally projections for the frozen spin method described in Sec.~\ref{sec:sen-frozen} are shown in purple. Also shown are existing limits from virtual contribution to muon g-2, as well as SN cooling, that effectively rule out a DM explanation to the g-2 anomaly from this operator. However, the frozen spin method could be sensitive to new parameter space. 

\subsubsection{$\partial_\alpha a^2 \, \bar{\mu} \gamma^\alpha \gamma_5 \mu$}

We could instead consider a CP violating operator
\begin{equation}
  \mathcal{L} \supset \partial_{\alpha} \lp
  \frac{a^2}{\Lambda^2} \rp \bar{\mu} \gamma_\alpha \gamma_5 \mu .
\end{equation}
This produces a RMRF precession analogous to Eqn.~\eqref{eq:wind-omega-rest}
\begin{align}
\label{eq:windsq-omega-rest}
  \vec{\omega}_r = \frac{1}{\Lambda^2} \; \vec{\nabla}\lp a^2 \rp.
\end{align}
In the lab frame we still have $a \approx a_0 \cos\lp m_a t \rp$, so that 
\begin{align} 
 a^2 \approx a_0^2 \lb \frac{1}{2} + \frac{1}{2} \cos\lp 2 m_a t \rp \rb.
\end{align}
Only the oscillatory term will contribute to Eqn.~\eqref{eq:windsq-omega-rest}, as it gets a spacial gradient upon boosting to the RMRF.
The frequency perturbation is 
\begin{align}
  \vec{\omega}_\x{dm} \approx 
  - \frac{2 \rho_\x{dm}}{m_a \, \Lambda} \, \vec{v} \sin\lp 2 \, \mdm t \rp
\end{align}
which is analogous to Eqn.~\eqref{eq:omdm-wind}.

\begin{figure}[htpb]
\centering
\includegraphics{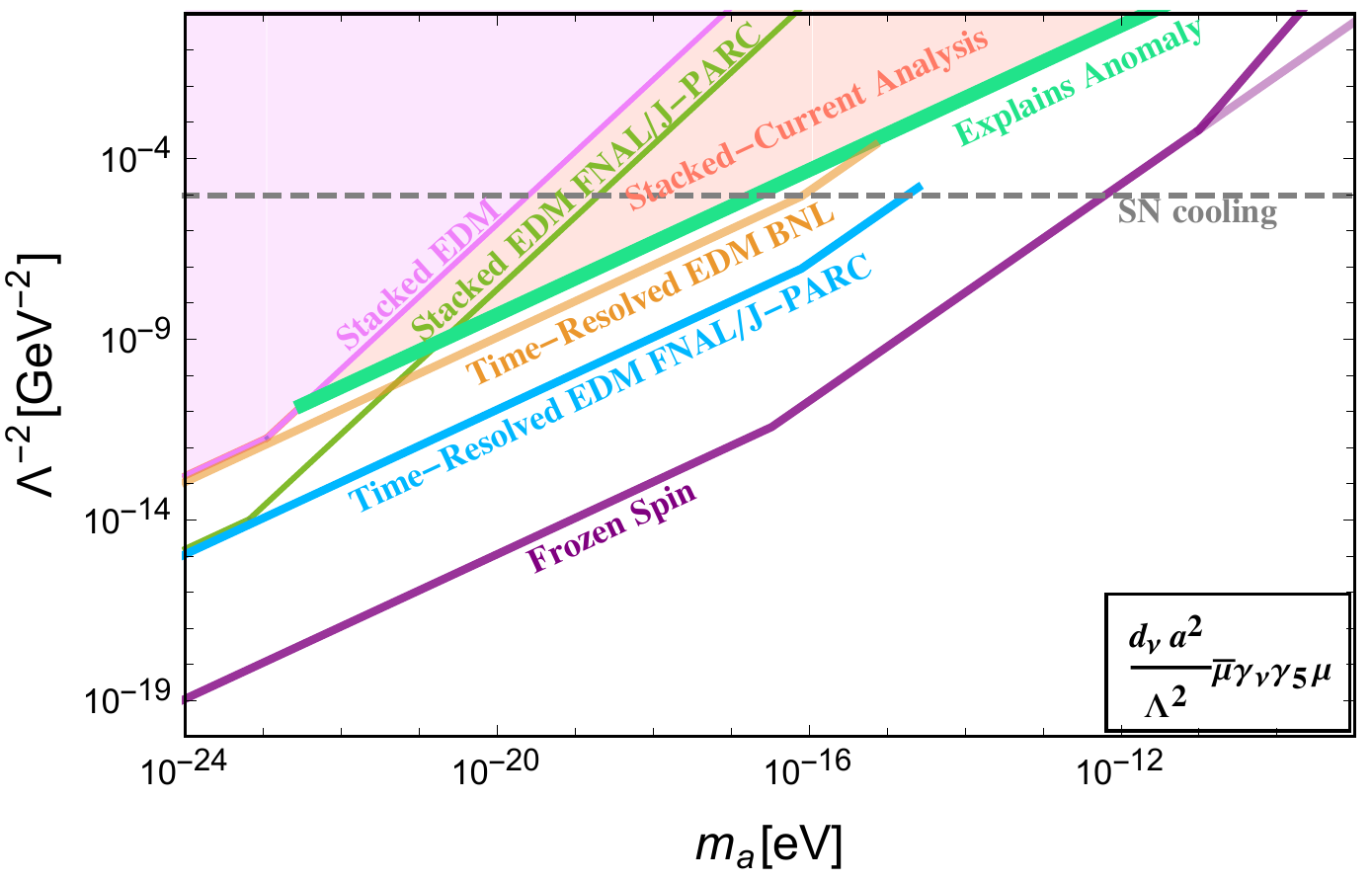}
\caption{Limits and projections for a pseudoscalar DM candidate $a$ with coupling $ \partial_{\alpha} \, 
  \frac{a^2}{\Lambda^2} \bar{\mu} \gamma_\alpha \gamma_5 \mu $ using the same color coding discussed in Fig.~\ref{fig3}.}
\label{fig4}
\end{figure}

Existing limits on $\Lambda$ now are weaker than in the linear case. The pseudoscalar must be pair produced inside stars and it occurs in two loops in vertex corrections to muon g-2. The same set of constraints as discussed in Section~\ref{sec:can-linear-wind} is applied to this operator and the results are plotted in Fig.~\ref{fig4}. 

\subsubsection{$a \bar{\mu}\sigma F\gamma_5 \mu$}
Finally, let us consider a pseudoscalar coupling only to muons via the operator
\begin{equation}
  \mathcal{L}\supset-i \frac{a}{2\Lambda^2} \bar{\mu}\sigma^{\alpha \beta} \gamma_5\mu F_{\alpha \beta} 
\end{equation}
 This generates a time dependent electric dipole moment for the muon given by,
\begin{equation}
  d_{\mu} = \frac{1}{\Lambda^2} \frac{\sqrt{2 \rho_{\rm dm}}} {m_a} 
  \cos \lp m_a t \rp 
\end{equation}
The contribution to the time-dependent precession frequency can be obtained from Eqn.~\eqref{eq:omega_a}. 
Ignoring the electric field, which is subdominant to $\vec{v}\times\vec{B}$~\cite{Bennett:2008dy}, we have 
\begin{equation}
\label{eq:omdm-edm}
\omega_{\rm dm}= 2 d_\mu \, \vec{v} \times \vec{B} = 
  \lp \vec{v} \times \vec{\omega}_{\rm sm} \rp \,
  \frac{m_{\mu}}{e a_{\mu} \Lambda^2} 
  \frac{\sqrt{2 \rho_{\rm dm}}} {m_a} \cos\lp m_a t\rp
\end{equation}

\begin{figure}[htpb]
\centering
\includegraphics{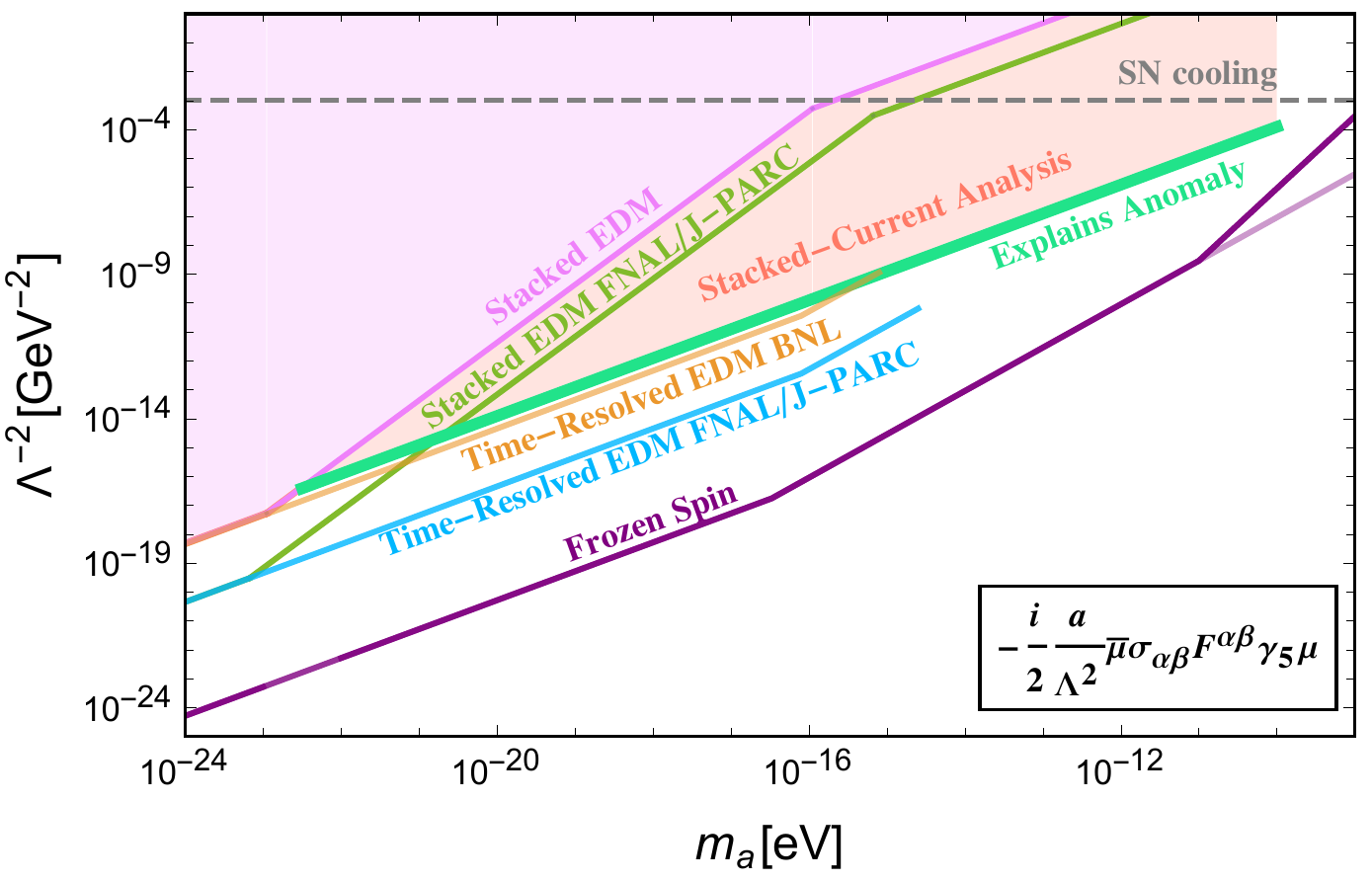}
\caption{Limits and projections for a pseudoscalar DM candidate $a$ with coupling $-i \frac{a}{2\Lambda^2} \bar{\mu}\sigma^{\alpha \beta} \gamma_5\mu F_{\alpha \beta} $ using the same color coding discussed in Fig.~\ref{fig3}}
\label{fig5}
\end{figure}

The DM perturbation is perpendicular to $\omega_{\rm sm}$ and is subject to the same limits and projections as considered in Section~\ref{sec:can-linear-wind}.
These are shown in Fig.~\ref{fig5}.  
The direct constraints on this operator from virtual contributions to muon g-2 are two-loop suppressed and are not shown.  This model does not possess a shift symmetry and constraints from radiatively generated couplings are discussed in Appendix.~\ref{looplevel}.

\subsection{Vectors}

\subsubsection{$L_\mu - L_\tau$}
\label{sec:can-vec}
\begin{figure}[htpb]
\centering
\includegraphics{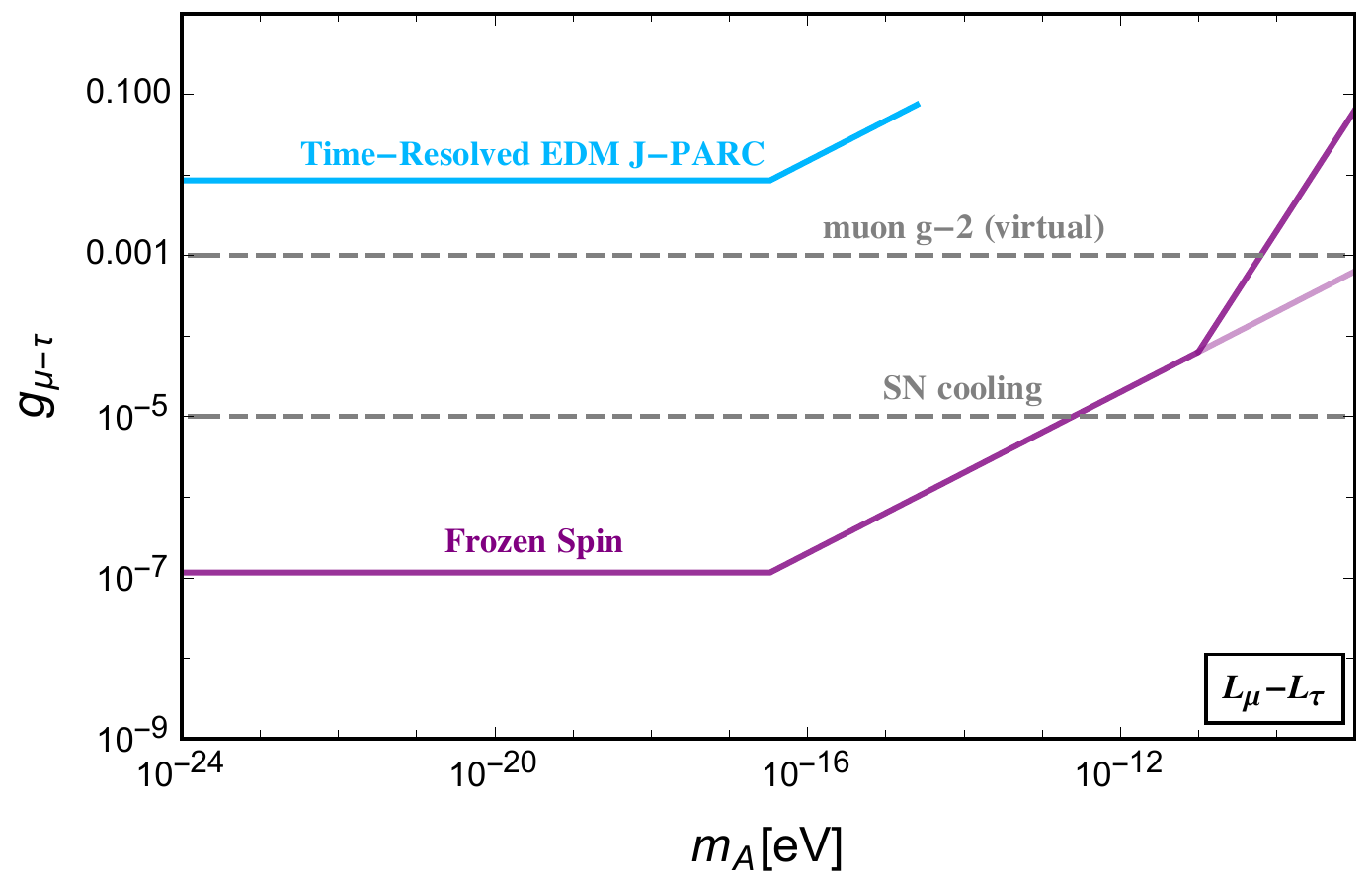}
\caption{Limits and projections for $L_{\mu}-L_{\tau}$ vector DM from current and future muon precession experiments are displayed. Projections in blue correspond to time-resolved analyses of g-2 data at the J-PARC experiment. Frozen spin experiments have a projected detection reach shown in dark (light) purple for a static (resonant) measurement. Shown in gray are constraints from virtual corrections to muon g-2 \cite{Altmannshofer:2014pba} and SN cooling \cite{Gninenko:1997iy}. Existing constraints from BNL and projections for Fermilab are suppressed due to their use of $\gamma=\gamma_\x{magic}$ and are not shown.  See Section~\ref{sec:can-vec} for details.}
\label{fig6}
\end{figure}

We consider an $L_\mu -L_\tau$ gauge boson as a vector DM candidate. With gauge coupling $\gmt$, this produces a local dark electric and magnetic field with magnitudes~\cite{Chaudhuri:2014dla}
\begin{align}
E_{\rm dm} = \sqrt{2 \rho_{\rm dm}} \cos\lp \mdm t + \alpha \rp \\
B_{\rm dm} = v_{\rm dm} \sqrt{2 \rho_{\rm dm}} \sin\lp \mdm t + \alpha \rp . 
\end{align} 
These fields apply both a spin torque and a force to muons, and yield a contribution to the RMRF precession frequency which has the same form as Eqn.~\eqref{eq:omega_a}
\begin{align}
\label{eq:omdm-vector}
 \vec{\omega}_\x{dm} = \frac{\gmt}{m_\mu} \lb a_\mu \vec{B}_\x{dm} -
  \lp a_\mu - \frac{1}{\gamma^2 - 1} \rp \lp \vec{v} \times \vec{E}_\x{dm} \rp -
  a_\mu \lp \frac{\gamma}{\gamma + 1} \rp \lp \vec{B}_\x{dm} \cdot \vec{v} \rp \vec{v} \rb,
\end{align}
where we have ignored any intrinsic muon EDM. 
It is helpful to decompose $\vec{B}_\x{dm}$ and $\vec{E}_\x{dm}$ into components along the vertical direction $B_{\x{dm},z}$, $E_{\x{dm},z}$, and components in the plane of the muon orbit $\vec{B}_{\x{dm}, \perp}$, $\vec{E}_{\x{dm}, \perp}$. 
We consider the effects of each of these four components in turn. 
\begin{enumerate}

\item
$E_{\x{dm},z}$ contributes to $\vec{\omdm}$ through the $\vec{v} \times \vec{E}_\x{dm}$ term of Eqn.~\eqref{eq:omdm-vector}. 
This term vanishes at BNL and Fermilab due to the use $\gamma_\x{magic}$ (see Section~\ref{sec:exp-deets}), but would otherwise yield
\begin{align}
\label{eq:omdm-Ez}
 \lp \frac{\omdm}{\omsm} \rp_{E_{\x{dm},z}} \approx
 \frac{\gmt}{e} \frac{1}{a_\mu \, \gamma^2} \frac{\sqrt{\rho_{\rm dm}}}{B}
 \approx 6 \cdot 10^{-4} \, \gmt \, \lp \frac{9}{\gamma^2} \rp \lp \frac{3~\x{T}}{B} \rp.
\end{align}
This is a harmonic, perpendicular perturbation which may be detected as described in Section~\ref{sec:sen-vc}. 
The projected detection reach of upcoming experiments is shown in Fig.~\ref{fig6} for the J-PARC experiment in blue and frozen spin experiments in purple.

\item 
$\vec{E}_{\x{dm}, \perp}$ yields a parallel perturbation if $\gamma \neq \gamma_\x{magic}$, in which case its amplitude is of the same order as Eqn.~\eqref{eq:omdm-Ez}. 
Since the direction of $\vec{E}_{\x{dm}, \perp}$ is constant in the lab frame and $\omdm \sim |\vec{v} \times \vec{E}_{\x{dm}, \perp}|$, the DM precession frequency now contains a product of two oscillations, one at frequency $\mdm$ and the other at the cyclotron frequency $\omega_C$. 
This yields two harmonic components with frequencies $\omega_C \pm \mdm$. 
$\omega_C$ is much faster than the frequency $\omsm$ at which $\vec{S}$ itself rotates, and so generically both of these components $\omega_C \pm \mdm$ are very rapid, which further suppressed the signal, as in Eqn.~\eqref{eq:Sp-parallel-harmonic}.
This suppression is removed in a narrow frequency interval around $\mdm \approx \omega_C$ in which case one of the components is nearly static. 
This signal is not presently observable in frozen spin experiments and may only be seen in the J-PARC total count, however even in this case the signal is to weak to be observed at the projected sensitivity. 

\item
$B_{\x{dm},z}$ is a harmonic, parallel perturbation with 
\begin{align}
\label{eq:omdm-Bz}
 \lp \frac{\omdm}{\omsm} \rp_{B_{\x{dm},z}} \approx
 \frac{\gmt}{e} \frac{\sqrt{\rho_{\rm dm}}}{B} v_\x{DM} 
 \approx 10^{-6} \, \gmt \,\lp \frac{3~\x{T}}{B} \rp,
\end{align}
which is too small to be observed by current sensitivity.. 
This is considerably weaker than the $E_{\x{dm},z}$ effect, as it is suppressed by both $v_\x{dm}$ and $a_\mu$. 

\item 
$\vec{B}_{\x{dm}, \perp}$ produces a perpendicular perturbation with an amplitude of the same order as that of $B_{\x{dm},z}$ in Eqn.~\eqref{eq:omdm-Bz}.
Similar to case of $\vec{E}_{\x{dm}, \perp}$, this produces perturbations which oscillate at frequencies $\omega_C \pm \mdm$. 
In this case, the two components of $\vec{\omega}_\x{dm}$ rotate in the RMRF.
By an argument analogous to Section~\ref{sec:trajectory-resonance-and-froze}, the vertical precession amplitude is then generally suppressed by an additional factor $\omsm/\omega_C \approx 10^{-3}$ which renders these perturbations unobservable with current sensitivity. 
This may be avoided in one of two narrow mass windows, either $|\omega_C - \mdm|/\omega_C \lesssim 10^{-3}$ in which case one of the components is slower then $\omsm$ and the signal follows Eqn.~\eqref{eq:SB-perp-wkb-harmoinc}, or $|\omega_C - \mdm - \omsm|/\omdm \ll 1$ which is the resonance regime discussed in Sections~\ref{sec:trajectory-resonance-and-froze} and~\ref{sec:sen-vc}.
We do not plot these cases as they are extremely narrow. 

\end{enumerate}

\subsection{Other dark relics}
The results presented thus far assume all of DM to be composed of the ultralight candidate under consideration. 
However, subcomponent dark matter may be easily tested as well --- the limits and projections presented here may be simply rescaled in the coupling plotted on the y-axis, either linearly or as the square-root of the DM fraction, depending on the candidate. 
For this reason we allow the mass range in our results to extend below the existing limit on fuzzy DM from dwarf galaxies~\cite{Safarzadeh:2019sre}.
In principle, these experiments are also sensitive to background fields that redshift differently than cold DM, such as dark radiation and dark energy. 
We leave a careful study of these candidates for future work. 

\section{Conclusion}
\label{sec:conclusion}
We have shown that experiments designed to measure the muon g-2 and EDM are uniquely sensitive to DM models that interact predominantly with muons. 
DM-induced variations in the properties of muons and DM-applied spin torques and forces on muons leads to time-dependent variations in the muon precession frequencies which are measured in these experiments. 
While an ultralight boson making up $\mathcal{O}(1)$ DM was the focus of this work, subcomponent DM, dark radiation, or even dark energy could in principle be observed through these precession experiments. 

Existing data from the muon g-2 experiments can be readily used to draw constraints on DM models that provide a perpendicular perturbation to the precession frequency, as these result in a net positive shift of the observed g-2 frequency. 
These models include the pseudoscalar wind couplings as well as pseudoscalar EDM-like couplings. 
Interestingly, a part of this parameter space also provides a unique explanation for the observed muon g-2 anomaly, which is distinct from solutions that invoke radiative corrections and which typically involve larger couplings between BSM and SM. 
This proposition could be tested by studying timing data of electron counts in existing EDM measurements at BNL or at the currently running Fermilab experiment. 
Dark matter models that contribute parallel perturbations are unlikely to explain the muon g-2 anomaly, but could also be tested using timing data. 
Lastly, vector DM produces an electric field whose effects are suppressed at BNL and Fermilab, which employ muons at the magic momentum. This effect could instead be discerned at the J-PARC experiment or with a frozen spin measurement, which uses slower muons. 
The most powerful detection opportunity available in the near future is the use of a time-resolved analysis in the frozen spin experiments proposed to measure the muon EDM, either in their intended static mode or repurposed as a resonant search.
Such an experiment can detect ultralight DM-muon interactions with unheralded sensitivity. 

\acknowledgments
We thank Jeff Dror, Patrick Fox, Roni Harnik, Jacob Leedom, Liang Li, Surjeet Rajendran, Paul Riggins, Tanner Trickle, and Vijay Narayan for useful discussions. H.R. is supported in part by the DOE under contract DE-AC02-05CH11231. 
\bibliography{bibliography}


\appendix

\section{Loop Level Constraints}
\label{looplevel}
In this section, we collect radiatively induced couplings and discuss constraints from such couplings on the operator considered as well as possible tunings. \\
\begin{center}{$\boldsymbol{1.~\phi \bar{\mu} \mu}$}\\
\end{center}
The operators induced at 1-loop by the Yukawa operator are:
\begin{equation}
\mathcal{L} \supset y \phi \left(\frac{2\alpha}{3m_\mu} F F + \frac{y_e y_\mu}{4\pi}  \bar{e}{e}+\frac{y_ny_\mu}{4\pi}  \bar{n}{n}\right) + \frac{y^2}{m_\mu} \phi^2 \left( \frac{y_e y_\mu}{4\pi}  \bar{e}{e}+\frac{y_n y_\mu}{4\pi}  \bar{n}{n}\right)
\end{equation}

\begin{figure}[htpb]
\centering
\includegraphics{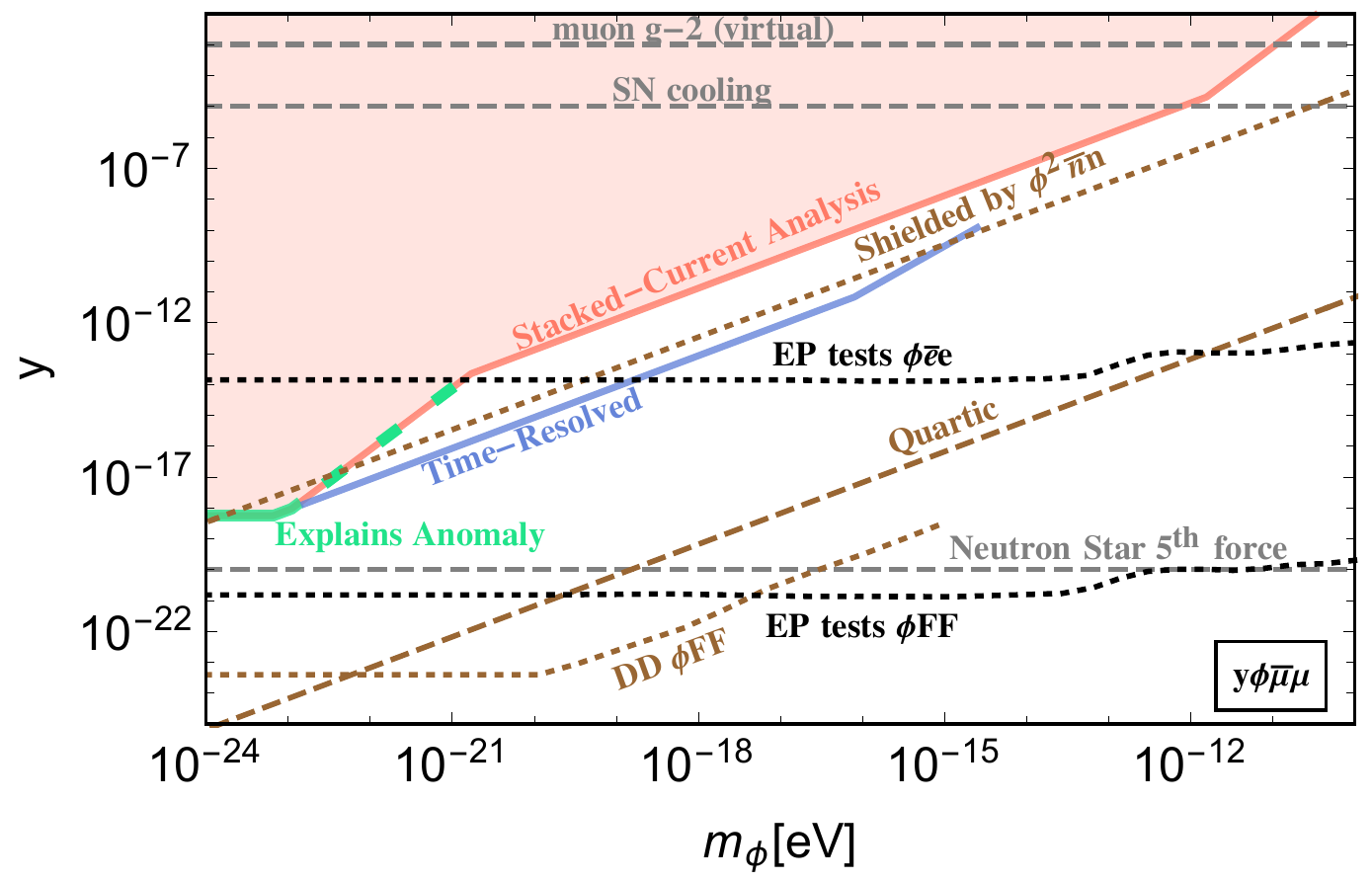}
\caption{Limits from radiatively induced operators on the Yukawa parameter space of Fig.~\ref{fig1}. EP tests (in black) and DD limits (in brown) from \cite{Arvanitaki:2016fyj} for photon and electron couplings are shown. Induced $\phi \bar{n}{n}$ can lead to shielding on earth above the relevant brown line. $\phi$ redshifts as DM only below the brown ``Quartic" line.}
\label{fig1A}
\end{figure}

Here $y_e$ is the SM electron Yukawa and $y_N$ is the effective Yukawa of the nucleon. 
The Yukawa type couplings, to a pair of photons, electrons and nucleons induced above have limits from stellar cooling, EP tests and also from atomic clocks if $\phi$ makes up all of dark matter. These are shown in Fig.~\ref{fig1A}. The $\phi^2 \bar{n}n$ and $\phi^2 \bar{e}{e}$ couplings induce a mass for the scalar in the presence of large SM number densities and can prevent the scalar from percolating into the earth. The estimate for this is,
\begin{equation}
\delta m_\phi^2 [\textrm{earth}]\sim \frac{y^2 }{m_\mu} \frac{y_n y_\mu}{4\pi} \frac{\rho_{\rm rock}}{m_n}\sim 6\times10^{-6} \textrm{eV}^2 y^2 \le m_\phi^2
\end{equation}
and is labeled in Fig.~\ref{fig1A} as ``shielded from $\phi^2 \bar{n}n$".
The Coleman Weinberg potential generates
\begin{equation}
\mathcal{L} \supset \frac{y^2}{4\pi^2} \Lambda_{\rm UV}^2 \phi^2+\frac{y^3}{24\pi^2}m_\mu \phi^3+\frac{y^4}{24\pi^2}\phi^4
\end{equation}
The mass term in the CW potential tells us how tuned the scalar is and  in general depends on the UV scale $\Lambda_{\rm UV}$. The quartic coupling generated needs to be small enough in order for $\phi$ to redshift like dark matter \cite{Arvanitaki:2014faa}. 
\begin{equation}
\lambda_{\rm eff} = \frac{y^4}{6 \pi^2}+ \frac{y^6}{36 \pi^4} \frac{m_\mu^2}{m_\phi^2} \le 3\times 10^{-79} \left(\frac{m_\phi}{10^{-18} \textrm{eV}}\right)^4.
\end{equation}
This is plotted as the ``Quartic" line in  Fig.~\ref{fig1A}.
These curves together show that the new Yukawa parameter space that can be probed by muon g-2 experiments is finely tuned and clever model building has to be performed in order to explain the absence of additional  operators that are severely constraining. 
\begin{center}
$2.~\boldsymbol{\phi^2 \bar{\mu} \mu}$\\
\end{center}
This radiatively generates,
\begin{equation}
\mathcal{L} \supset  \frac{\phi^2}{\Lambda} \left(\frac{2\alpha}{3m_\mu} F F + \frac{y_e y_\mu}{4\pi}  \bar{e}{e}+\frac{y_ny_\mu}{4\pi}  \bar{n}{n}\right)+\frac{\phi^4}{16\pi^2}\frac{m_\mu^2}{\Lambda^2}
\end{equation}

\begin{figure}[htpb]
\centering
\includegraphics{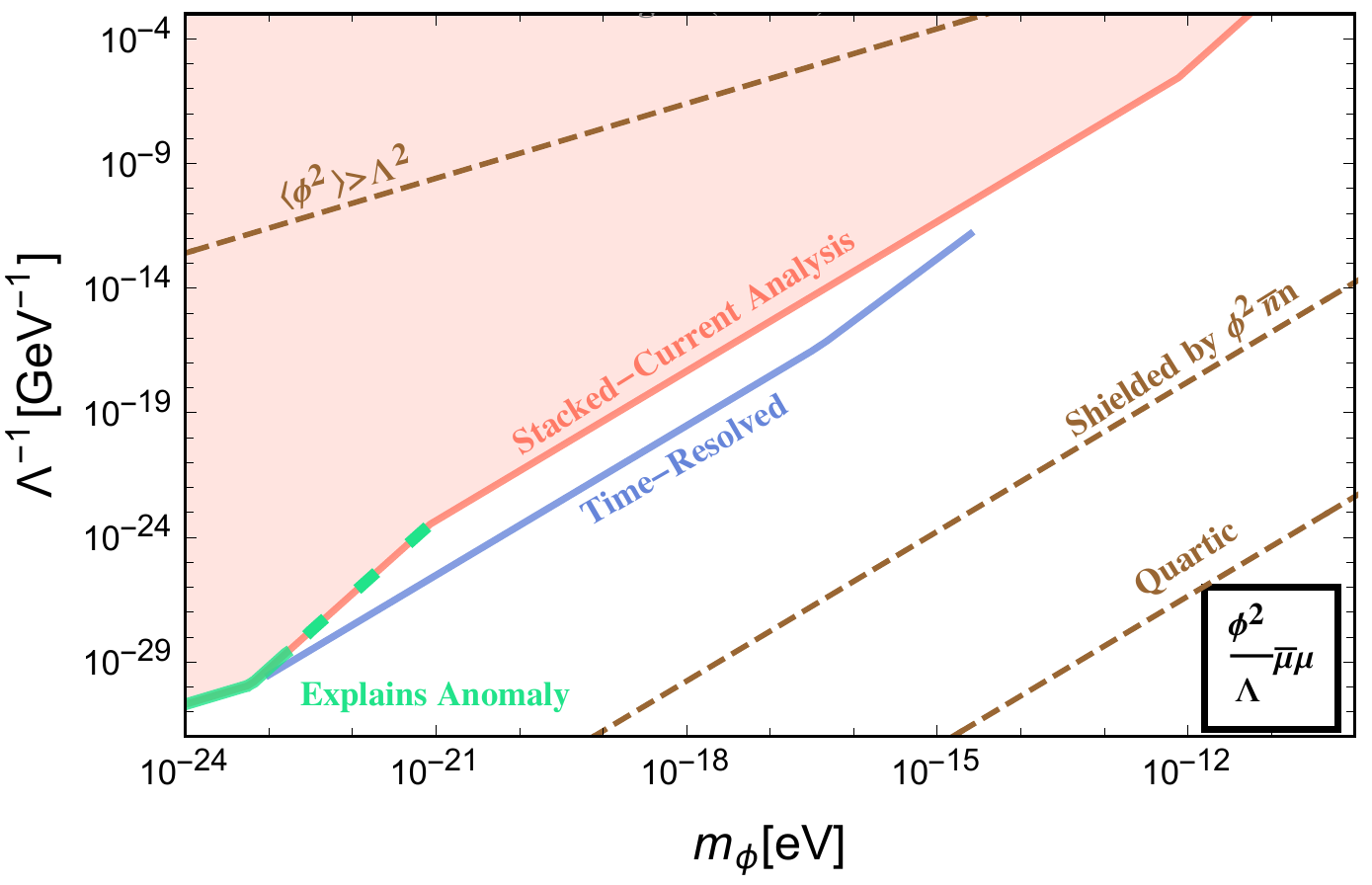}
\caption{Limits from radiatively induced operators on the $\frac{\phi^2}{\Lambda}\bar{\mu}\mu$ parameter space of Fig.~\ref{fig2}. Quartic and shielding limits are similar to Fig.~\ref{fig1A}. Also shown is the region where $\langle \phi^2 \rangle \ge \Lambda$ where the EFT might not be well-defined.}
\label{fig2A}
\end{figure}

\begin{figure}[htpb]
\centering
\includegraphics{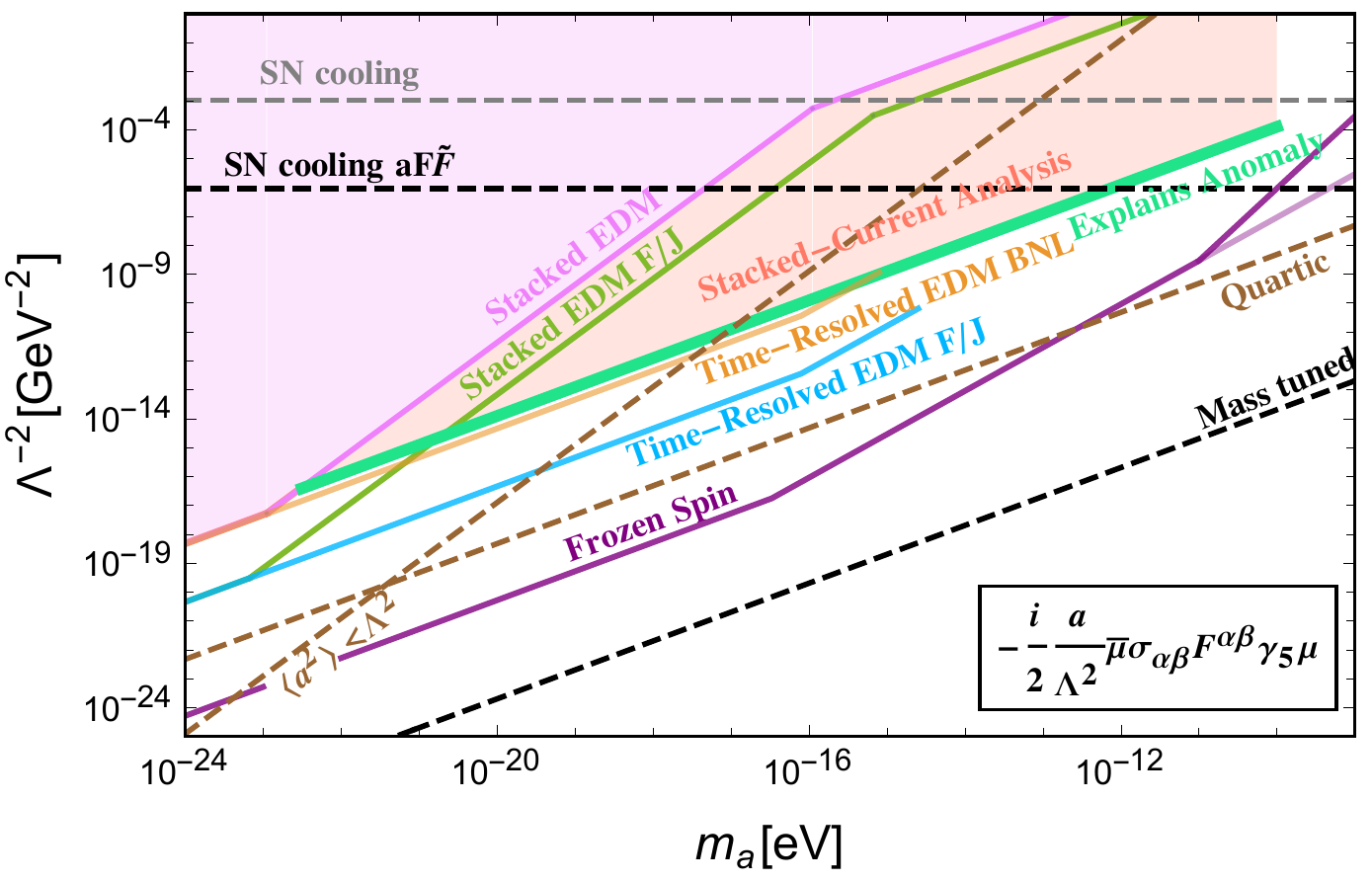}
\caption{Limits from radiatively induced operators on the $a \bar{\mu}\sigma F\gamma_5 \mu$ parameter space of Fig.~\ref{fig5}. Supernova limits from the induced coupling to photons is shown in black. Quartic limits in brown are similar to Fig.~\ref{fig1A}. Also shown is the brown line above which $\langle a^2 \rangle \ge \Lambda$ where the EFT might not be well-defined. The region below the black tuning line corresponds to natural parameter space where the coupling is weak enough to accommodate light masses naturally.}
\label{fig3A}
\end{figure}

Just like the Yukawa case, $\phi^2 \bar{n}n$ and $\phi^2 \bar{e}{e}$ can prevent scalar from percolating into the earth. This is given by, 
\begin{equation}
\delta m_\phi^2 [\textrm{earth}]\sim \frac{1}{\Lambda} \frac{y_n y_\mu}{4\pi} \frac{\rho_{\rm rock}}{m_n}\sim 6\times10^{-10} \textrm{eV}^2 \frac{\rm TeV}{\Lambda} \le m_\phi^2
\end{equation}
Requiring small enough quartic gives,
\begin{equation}
\lambda_{\rm eff} = \frac{\Lambda_c^2}{16 \pi^2\Lambda^2} \le 3\times 10^{-79} \left(\frac{m_\phi}{10^{-18} \textrm{eV}}\right)^4
\end{equation}
Finally, depending on the details of UV physics, the EFT is safe only for field values well below the cutoff scale, i.e. $\phi_{\rm DM} \le \Lambda$. 
These constraints are plotted in Fig.~\ref{fig2A}. 
\\
\begin{center}
$3.~\boldsymbol{a \bar{\mu}\sigma F\gamma_5 \mu}$\\
\end{center}
At one loop, the EDM operator generates
\begin{equation}
\mathcal{L} \supset \frac{e}{4\pi^2}\frac{m_\mu}{\Lambda^2}(a F\tilde{F})+ \frac{e}{4\pi^2}\frac{m_\mu}{\Lambda^2}(\partial_\alpha a \mu \gamma_\alpha \gamma_5\bar{\mu})
\end{equation}
The first operator leads to $\Lambda \gtrsim 3$ TeV as shown in Fig.~\ref{fig3A}. The only rigorous limit on the second operator comes from muon g-2, and this should be sub-leading. 

At 2-loop and 3-loop, the self-interactions are
\begin{equation}
\mathcal{L} \supset \frac{1}{(4\pi)^4} \frac{m_\mu^6}{\Lambda^4} a^2+ \frac{1}{(4\pi)^6} \frac{m_\mu^8}{\Lambda^8} a^4
\end{equation}

This roughly corresponds to tuned masses when,
\begin{equation}
\delta m_a\sim 0.01 \textrm{eV} \left(\frac{\rm TeV}{\Lambda}\right)^2 \gtrsim m_a
\end{equation}
Constraining the quartic for $a$ to redshift like DM gives, 
\begin{equation}
\lambda_a \sim 10^{-36}\left(\frac{\rm TeV}{\Lambda}\right)^8 \le 3\times 10^{-79} \left(\frac{m_\phi}{10^{-18} \textrm{eV}}\right)^4
\end{equation}
These tuning lines are shown in Fig.~\ref{fig3A}.

Finally, $a^2 \bar{n}n$ and $a^2 \bar{e}{e}$ can prevent percolation into the earth. These radiatively generated couplings are:
\begin{equation}
\mathcal{L}=\frac{1}{(4\pi)^2}\frac{m_\mu^3 a^2}{\Lambda^4} \left( \frac{y_e y_\mu}{4\pi}  \bar{e}{e}+\frac{y_n y_\mu}{4\pi} \bar{n}n \right)
\end{equation}
which give a correction,
\begin{equation}
\delta m_a^2 [\textrm{earth}]\sim \frac{1}{(4\pi)^2}\frac{m_\mu^3}{\Lambda^4} \frac{y_n y_\mu}{4\pi} \frac{\rho_{\rm rock}}{m_n}\sim 3\times10^{-23} \textrm{eV}^2 \left(\frac{\rm GeV}{\Lambda}\right)^4 \le m_a^2
\end{equation}
But this is subleading and not shown in the plot. 
\end{document}